\documentclass[journal]{IEEEtran}

\usepackage{amssymb, graphicx, booktabs, longtable}
\usepackage{cite}
\usepackage{amsmath}
\usepackage{color}
\usepackage{multirow}
\usepackage{changes} 
\usepackage{todonotes}

\ifCLASSINFOpdf
\else
\fi
\begin{document}
%
\title{ A Latent Encoder Coupled Generative Adversarial Network (LE-GAN) for Efficient Hyperspectral Image Super-resolution}%
%
%

\author{Yue~Shi, Liangxiu~Han*, Lianghao Han, Sheng~Chang, Tongle Hu, Darren Dancey
        
\thanks{Yue Shi, Liangxiu Han, Darren Dancey are with Department of Computing and Mathematics, Faculty of Science and Engineering, Manchester Metropolitan University, Manchester M1 5GD, UK.}
\thanks{Lianghao Han is with Department of Computer Science, Brunel University, UB8 3PH, UK}
\thanks{ Sheng Chang is with State Key Laboratory of Remote Sensing Science, Aerospace Information Research Institute, Chinese
Academy of Sciences, Beijing 100101, China.}
\thanks{Tongle Hu is College of Plant Protection, Hebei Agriculture University, Baoding 070001, China}
\thanks {Corresponding author: L. Han (e-mail: l.han@mmu.ac.uk)}
}

%
%

\markboth{Journal of \LaTeX\ Class Files,~Vol.~13, No.~9, September~2014}%
{Shell \MakeLowercase{\textit{et al.}}: Bare Demo of IEEEtran.cls for Journals}
%



\maketitle

\begin{abstract}
Realistic hyperspectral image (HSI) super-resolution (SR) techniques aim to generate a high-resolution (HR) HSI with higher spectral and spatial fidelity from its low-resolution (LR) counterpart. The generative adversarial network (GAN) has proven to be an effective deep learning framework for image super-resolution. However, the optimisation process of existing GAN-based models frequently suffers from the problem of mode collapse, leading to the limited capacity of spectral-spatial invariant reconstruction. This may cause the spectral-spatial distortion on the generated HSI, especially with a large upscaling factor. To alleviate the problem of mode collapse, this work has proposed a novel GAN model coupled with a latent encoder (LE-GAN), which can map the generated spectral-spatial features from the image space to the latent space and produce a coupling component to regularise the generated samples. Essentially, we treat an HSI as a high-dimensional manifold embedded in a latent space. Thus, the optimisation of GAN models is converted to the problem of learning the distributions of high-resolution HSI samples in the latent space, making the distributions of the generated super-resolution HSIs closer to those of their original high-resolution counterparts.  
We have conducted experimental evaluations on the model performance of super-resolution and its capability in alleviating mode collapse. The proposed approach has been tested and validated based on two real HSI datasets with different sensors (i.e. AVIRIS and UHD-185) for various upscaling factors (i.e. $\times 2$, $\times 4$, $\times 8$) and added noise levels (i.e. $\infty$ db, $40$ db, $80$ db), and compared with the state-of-the-art super-resolution models (i.e. HyCoNet, LTTR, BAGAN, SR- GAN, WGAN). 
Experimental results show that the proposed model outperforms the competitors on the super-resolution quality, robustness, and alleviation of mode collapse. The proposed approach is able to capture spectral and spatial details and generate more faithful samples than its competitors. It has also been found that the proposed model is more robust to noise and less sensitive to the upscaling factor and has been proven to be effective in improving the convergence of the generator and the spectral-spatial fidelity of the super-resolution HSIs.

\end{abstract}


%
\IEEEpeerreviewmaketitle

\section{Introduction}
\label{sec:1}
\IEEEPARstart{T}{he} hyperspectral image (HSI) has been widely used in extensive earth observation applications because of the rich information in its abundant spectral bands. However, due to the cost and hardware limitations of imaging systems, the spatial resolution of HSI decreases when the numerous spectral signals are collected simultaneously \cite{deng2021m2h, lanaras2018super, palsson2017multispectral}. Due to this drawback, the HSI does not always meet the demands for high-accurate earth observation tasks. The HSI super-resolution aiming to estimate a high-resolution (HR) image from a single low-resolution (LR) counterpart is one of promising solutions. Currently, there are mainly two different approaches for HSI super-resolution: 1) the HSI fusion with the HR auxiliary image (e.g. panchromatic image) and 2) the single HSI super-resolution without any auxiliary information. Generally, the image fusion approach implements the super-resolution using filter-based approaches through integrating the high-frequency details of HR auxiliary image into the target LR HSI \cite{yokoya2017hyperspectral, ghamisi2019multisource}, such as component substitution \cite{qu2017hyperspectral, khademi2017multi}, spectral unmixing \cite{brezini2021hypersharpening, lanaras2017hyperspectral}, and Bayesian probability \cite{benediktsson2018remote, wang2018high}. However, this method highly relies on the high-quality auxiliary image with high imaging cost, which limits its practical applications. In contrast, single HSI super-resolution does not need any other prior or auxiliary information, which has greater practical feasibility.  \par

In recent years, the single HSI super-resolution technologies have attracted increasing attention in remotely sensed data enhancement \cite{RN17}. Particularly, Deep Learning (DL)-based single image super-resolution (SISR) methods have achieved significant performance improvement \cite{LN02}. The first DL-based method for single image super-resolution was proposed by Dong \textit{et al.} \cite{LN01}, named as the super-resolution convolutional neural network (SRCNN). To recover the finer texture details from low-resolution HSIs with large upscaling factors, Ledig \textit{et al.} \cite{RN2} proposed a super-resolution generative adversarial network (SRGAN) by introducing a generative adversarial network (GAN). After that, various GAN-based deep learning models have been developed and proven to be effective in improving the quality of image super-resolution \cite{RN8, RN3, RN10, RN6}. \par

\begin{figure}[h]   
    \centering  
    \includegraphics[width=3.6in]{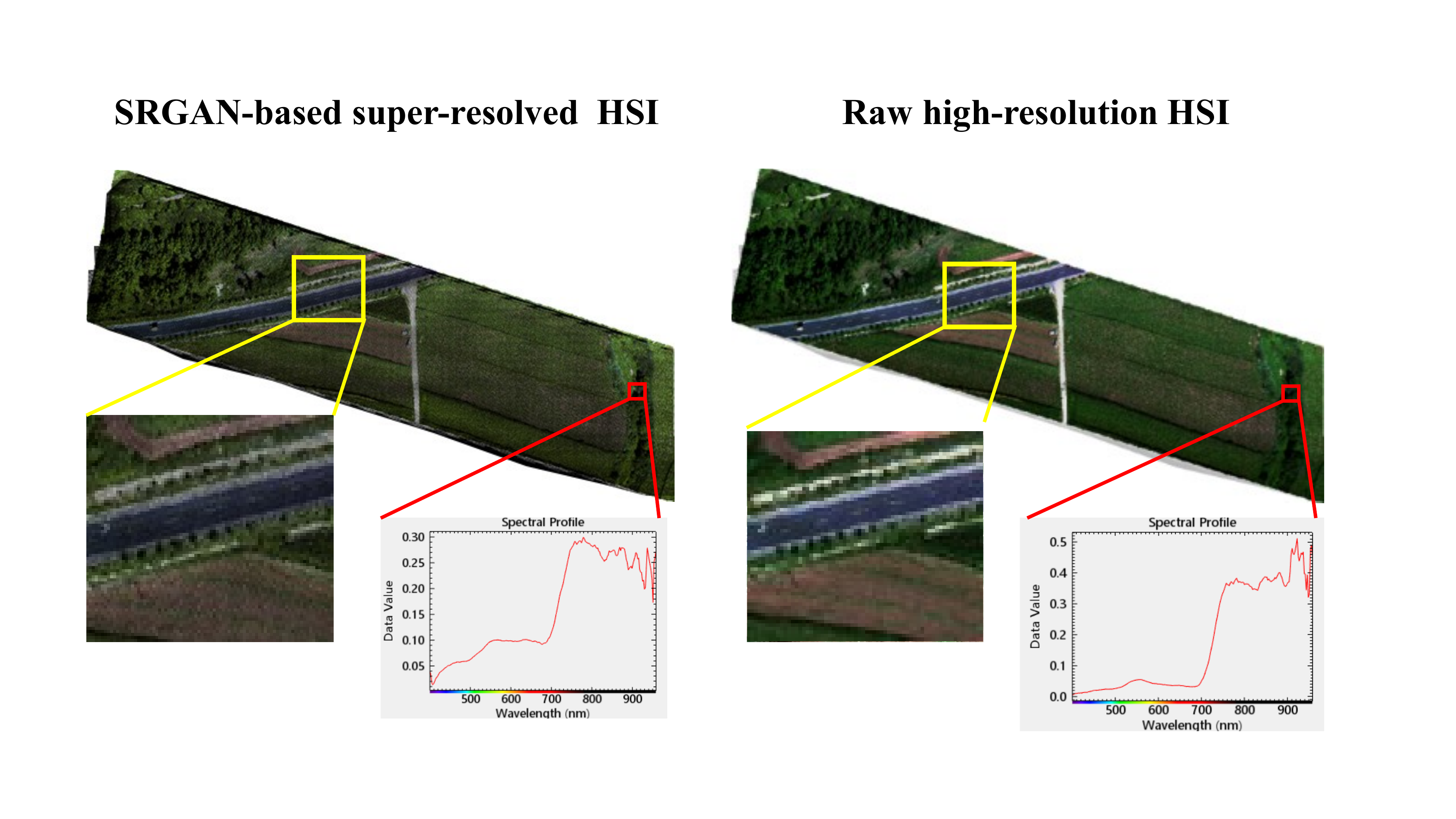}   
    \caption{A comparison between a raw high-resolution (right) HSI and its (8 $\times$) super-resolution HSI (left) counterpart generated by the SRGAN model \cite{RN2}. The red frames show the spectral distortion occurs in the learning process, and the yellow frames reveal the loss of spatial details in the super-resolution HSI.}    
    \label{fig:1}  
\end{figure}

However, existing GAN-based super-resolution approaches mainly focused on RGB images, in which the reflectance radiance characteristics between the neighbouring spectral channels were not considered in the model training processes. Therefore, using these models for HSI super-resolution directly will lead to the absence of spectral-spatial details in the generated images. For example, Fig.\ref{fig:1} shows a comparison between an original high-resolution HSI and its super-resolution HSI counterpart generated from the SRGAN model \cite{LN02}. Obvious spectral-spatial distortions can be observed on the generated super-resolution HSI (see the red and yellow frames in Fig.\ref{fig:1}). 
Mathematically, recovering spectral-spatial details in super-resolution HSI is an under-determined inverse problem in which a large number of plausible details in the high-resolution image need to be characterised from low-resolution information. The complexity of this under-determined issue will exponentially increase with the increased upscaling factor. With high upscaling factors (e.g. higher than 8 $\times$ ), the spectral-spatial details of generated super-resolution HSIs could be distorted.   \par

The potential reason behind the spectral-spatial distortion is due to mode collapse in the optimisation process of GANs \cite{xiong2020improved, li2020tackling}, in which GAN models get stuck in a local minimum and only learn limited modes of data distributions. Some studies have attempted to address mode collapse in GAN models. For instance, Hou \textit{et al.} \cite{hou2020unsupervised} improved the diversity of the generator in GAN models and attempted to avoid the mode collapse by adding a reverse generating module and an adaptive domain distance measurement module into the GAN framework. Their findings illustrated that these approaches facilitated solving the insufficient diversity of GAN models in remote sensing image super-resolution. Ma \textit{et al.} \cite{ma2019super} introduced a memory mechanism into GAN models to save feedforward features and extract local dense features between convolutional layers, which showed some effectiveness in increasing spatial details during the reconstruction procedure. \par

To benefit the remarkable super-resolution performance from GAN-based models and address the spectral-spatial distortions in HSI super-resolution, in this study, we proposed a novel latent encoder coupled GAN architecture. We treated an HSI as a high-dimensional manifold embedded in a higher dimensional ambient latent space. The optimisation of GAN models was converted to a problem of learning the feature distributions of high-resolution HSIs in the latent space, making the spectral-spatial feature distributions of generated super-resolution HSIs close to those of their original high-resolution counterparts. Our contributions included: \par

1) A novel GAN-based framework has been proposed to improve HSI super-resolution quality. The improvement was achieved from two aspects. Firstly, for improving the spectral-spatial fidelity, a short-term spectral-spatial relationship window (STSSRW) mechanism has been introduced to the generator in order to facilitate spectral-spatial consistency between the generated super-resolution and real high-resolution HSIs in the training process. Secondly, for alleviating the spectral-spatial distortion, a latent encoder has been introduced into the GAN framework as an extra module to make the generator do a better estimation on local spectral-spatial invariance in the latent space.   \par

2) A spectral-spatial realistic perceptual (SSRP) loss has been proposed to guide the optimisation of the under-determined inverse problem and alleviate spectral-spatial mode collapse issues occurred in the HSI super-resolution process, and benefit on retrieving high-quality spectral-spatial details in the super-resolution HSI, especially for high upscaling factors (e.g. 8$\times$). The loss function, SSRP, was able to enforce spectral-spatial invariance in the end-to-end learning process and made the generated super-resolution features closer to the manifold neighbourhood of the targeted high-resolution HSI. \par

The rest of this work is organised as follows: Section 2 introduces related works on existing GANs-based methods for HSI super resolution tasks; Section 3 details the proposed approach; Section 4 presents experimental evaluation results; Section 5 concludes the work.  \par

\section{Related work}
A traditional GAN-based super-resolution model contains two neural networks, a generator producing sample images from low-resolution images and a discriminator distinguishing real and generated images \cite{RN18}. The generator and discriminator are trained in an adversarial fashion to reach a Nash equilibrium in which the generated super-resolution samples become indistinguishable from real high-resolution samples.\par
Focusing on the spectral and spatial characteristics of HSI data, various adversarial strategies were proposed to improve the GAN performance on HSI super-resolution tasks \cite{RN20}. For example, Zhu \textit{et al} \cite{RN21} proposed a 3D-GAN to improve the generalisation capability of the discriminator in spectral and spatial feature classification with limited ground truth HSI data. Jiang \textit{et al} \cite{RN13} designed a spectral and spatial block inserted before the GAN generator in order to extract high-frequency spectral-spatial details for reconstructing super-resolution HSI data. \par

Some methods for improving the overall visual quality of generated HSIs were also proposed through constructing a reliable mapping function between LR and HR HSI pairs. For example, Li \textit{et al} \cite{RN22} proposed a GAN-based model for multi-temporal HSI data enhancement. In their model, a 3DCNN based upscaling block was used to collect more texture information in the upscaling process. Huang \textit{et al} \cite{RN23} integrated the residual learning based gradient features between an LR and HR HSI pair with a mapping function in the GAN model, and achieved the HSI super-resolution with an improved spectral and spatial fidelity.\par

The performance of a GAN-based model mainly depends on its generator, its discriminator and loss functions. Therefore, existing studies in improving GAN-based models for HSI resolution focused on their design and optimisation.

\subsection{Design of the generator and the discriminator}
In the generator where LR data are upscaled to a desired size, upscaling filters are the most important components that influence the performance of the generator in term of accuracy and speed \cite{RN25, RN24}. Ledig \textit{et al.} \cite{RN2} employed a deep ResNet with a skip-connection in the generator to produce super-resolution images with $\times 4$ upscaling factor. Jiang \textit{et al.} \cite{RN8} proposed an edge-enhancement GAN generator in which a group of dense layers were introduced into the generator in order to capture intermediate high-frequency features and recover the high-frequency edge details of HSI data.

In regard to the discriminator, it was found that a deeper network architecture had greater potential in discriminating real images from generated ones \cite{RN26, RN2}. For example, Rangneka \textit{et al.} \cite{RN27} proposed a GAN-based deep convolutional neural network with seven convolutional layers in the discriminator for aerial spectral super-resolution. Arun \textit{et al.} \cite{RN28} used six 3D convolutional filters and three deconvolution filters in the discriminator to discriminate the spectral-spatial features of real HR HSIs from the generated counterparts.

In the design of the generator and the discriminator, the computational cost needs to be considered. The upscaling process in the generator can significantly increase the computational cost at the scale of $n^2$ times for an the upscaling factor of $n$. Meanwhile, the deep learning-based discriminator always requires a large amount of computational time and memory for extracting and discriminating the high-dimensional non-linear mapping features of input data. More efficient generator and discriminator are required for fast and accurate HSI super-resolution.

\subsection{Design of loss functions}
The loss function plays a very important role in optimising the performance of GAN models \cite{RN30, RN29}. In the traditional GAN model, the generator and the discriminator are trained simultaneously to find a Nash equilibrium to a two-player non-cooperative game. A min-max loss function is used, it is equivalent to minimising Jensen-Shannon (JS) divergence between the distributions of generative data and real samples when the discriminator is optimal. However, the GAN training is hard, and can be slow and unstable. There are some issues in the original GAN model, such as hard to achieve Nash Equilibrium, the problem of low dimensional supports of sample distributions and mode collapse \cite{liu2020posegan, zhang2020ndvi}. To facilitate the training stability and address mode collapse problems in the original GAN model, several improved adversarial loss functions were developed, which can be divided into three categories: 1) the pixel-wised loss, 2) the perceptual loss, and 3) the probabilistic latent space loss. \par

In the first category, the pixel-wised mean squared error (MSE) loss is commonly used for measuring the discriminative difference between real and generated data in GAN models \cite{RN2}. However, the MSE has some issues, such as the loss of high-frequency details, the over-smoothing problem, and the sparsity of weights \cite{jiang2019respiratory, yarlagadda2018satellite, bhattacharjee2018posix}. Some studies have attempted to solve these issues. Chen \textit{et al.} \cite{chen2018attention} introduced a sparse MSE function into the GAN model in order to measure the high-frequency information in the spatial attention maps of images, their results showed that the GAN with the sparse MSE loss was able to provide more viable segmentation annotations of images. Zhang \textit{et al.} \cite{zhang2020supervised} emphasised that the MSE-loss always led to an over-smoothing issue in the GAN optimisation process. Therefore, they introduced a supervised identity-based loss functions to measure the semantic differences between pixels in the GAN model. Lei \textit{et al.} \cite{lei2019gcn} attempted to solve the issue of sparsity of the pixel-wised weights in the GAN model, and proposed two additional metrics, the edge-wise KL-divergence and the mismatch rate, for measuring the sparsity of pixel-wised weights and the wide-value-range property of edge weights.   \par

In the second category, existing studies used different perceptual losses to balance the perceptual similarity, based on high-level features and pixel similarity. Cha \textit{et al.} \cite{cha2017adversarial} proposed three perceptual loss functions in order to enforce the perceptual similarity between real and generated images, these functions achieved improved performance on generating high-resolution image with GAN models. Luo \textit{et al.} \cite{luo2018bi} introduced a novel perceptual loss into the GAN based SR model, named as Bi-branch GANs with soft-thresholding (Bi-GANs-ST), to improve the objective performance. Blau \textit{et al.} \cite{blau2018perception} proposed a perceptual-distortion loss function in which the generative and perceptual quality of GAN models were jointly quantified. Rad \textit{et al.} \cite{rad2019srobb} proposed a pixel-wise segmentation annotation to optimise the perceptual loss in a more objective way. Their model achieved a great performance in finding targeted perceptual features. \par

In the third category, Bojanowski \textit{et al.} \cite{RN11} investigated the effectiveness of the latent optimisation for GAN models, and proposed a Generative Latent Optimisation (GLO) strategy for mapping the learnable noise vector to the generated images by minimising a simple reconstruction loss. Compared to a classical GAN, the GLO obtained competitive results but without using the adversarial optimisation scheme which was sensitive to random initialisation, model architectures, and the choice of hyper-parameter settings. Training a stable GAN model is challenging. Therefore, Wasserstein GAN \cite{LN04} was proposed to improve the stability of learning and reduce mode collapse. The WGAN replaced the discriminator model with a critic which scored the realness of a given image in the probabilistic latent space and was trained using Wasserstein loss. Rather than discriminating between real and generated images (i.e. the probability of a generated image being real), the critic maximises the difference between its prediction for real images and generated images (i.e. predict a "realness" score of a generative image). Gulrajani \textit{et al.} \cite{RN32} further improved the WGAN training by adding a regularisation term penalising the deviation of the critic's gradient norm with regard to the input, and the model was named as (WGAN-GP).


\section{The proposed LE-GAN for single HSI super-resolution}
To address the challenge of spectral-spatial distortions caused by mode collapse during the optimisation process, we proposed a novel GAN model coupled with a latent encoder, named as LE-GAN. In the proposed framework, the optimised generator and discriminator were designed to improve the super-resolution performance and reduce the computational complexity. Inspired by the encoder coupled GAN \cite{RN11}, we developed a latent encoder embedded into our GAN framework to facilitate the generator to achieve a better approximation on feature maps, in order to generate ideal super-resolution results. In addition, we designed a spectral-spatial realistic perceptual (SSRP) loss function in order to optimise the under-determined inverse problem by providing a trade-off between aligning the distributions of generated super-resolution and targeted high-resolution HSIs and increasing the spectral-spatial consistency between them.\par

\subsection{Model architecture}
\label{sec:March}

\begin{figure}[!t]   
    \centering  
    \includegraphics[width=3.6in]{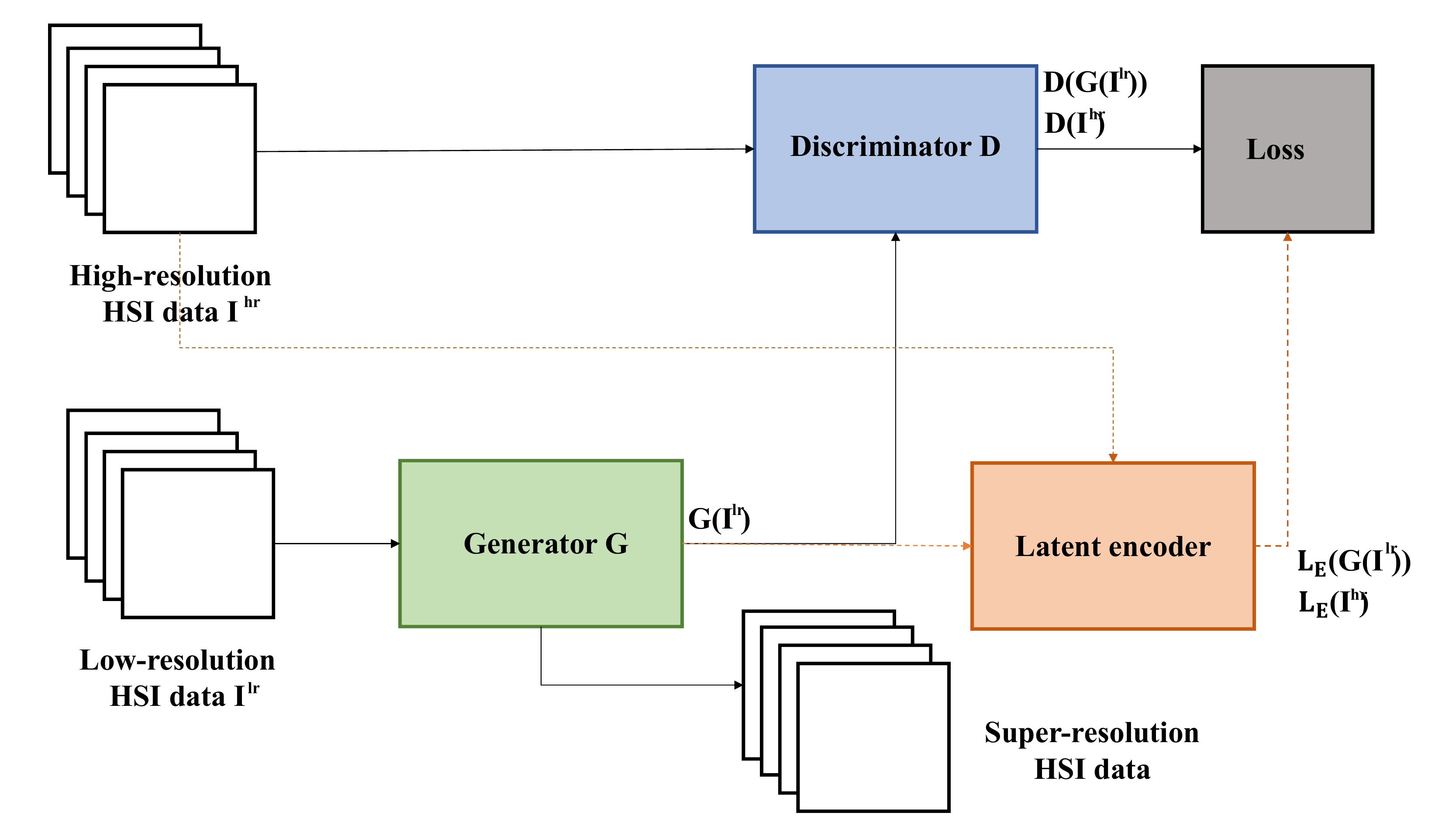}   
    \caption{The architecture of the proposed model: the output from the encoder is used to regularise the loss function}    
    \label{fig:Networkarchitecturte}  
\end{figure}

We have made two major changes to the traditional GAN framework: 1) proposed an improved generator, denoted as $G$, with a simplified ResNet structure, and 2) introduced a latent encoder, denoted as $L_E$, into the GAN framework. The network architecture is shown in Fig.\ref{fig:Networkarchitecturte}. It consists of a generator, a discriminator and an encoder. 

\subsubsection{The architecture of the generator model $G$}

\begin{figure}[!t]   
    \centering  
    \includegraphics[width=3.6in]{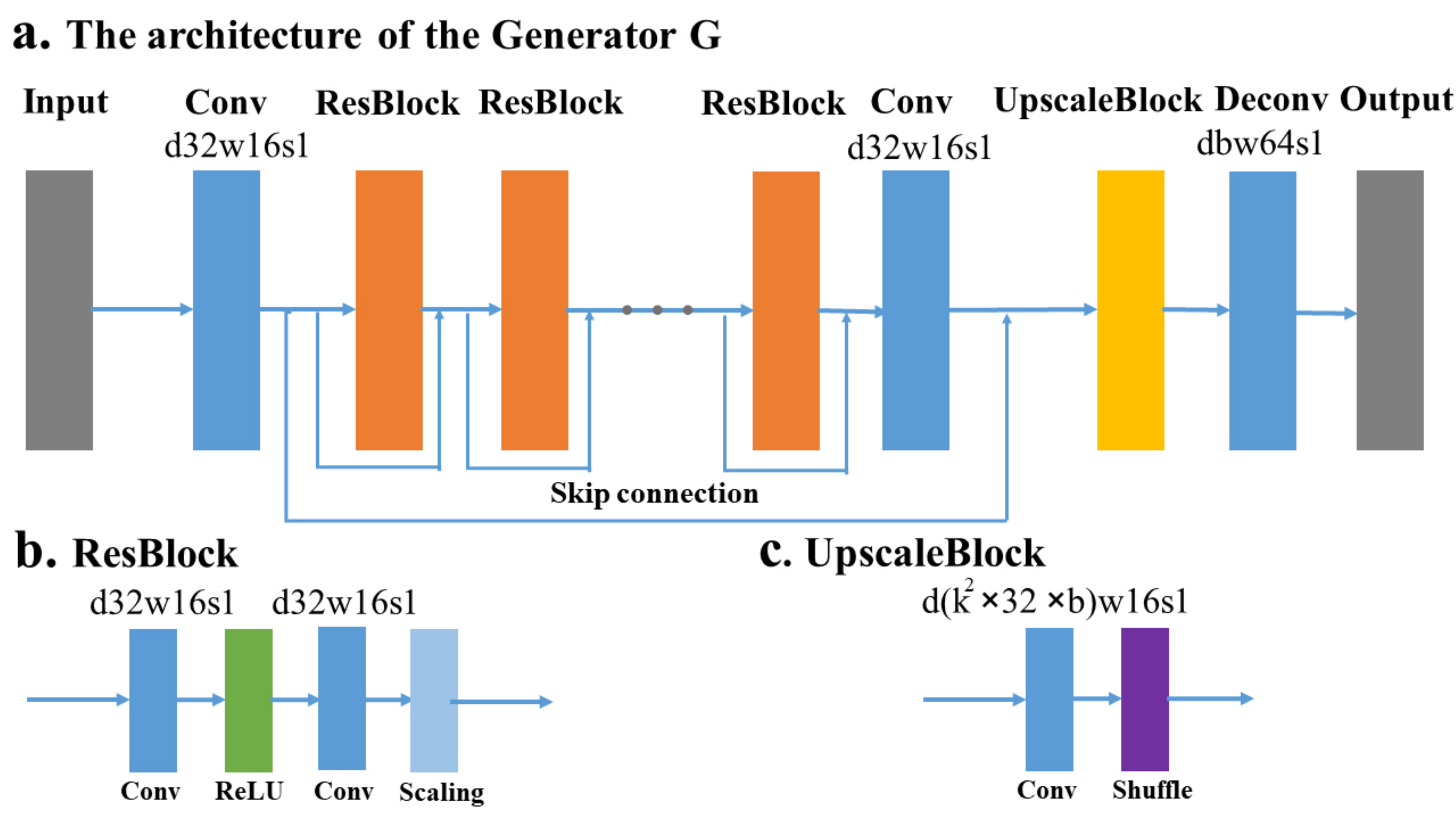}   
    \caption{The architectures of (a) the generator G, (b) the ResBlock component and (c) the UpscaleBlock component. Note: d,w,s are the kernel depth, the kernel width and the stride for a convolutional layer.}    
    \label{fig:G}  
\end{figure}

To improve the spectral-spatial reconstruction quality with low distortion and reduce the computational complexity, a short-term spectral-spatial relationship window (STSSRW) derived model was proposed, denoted as $G$ in our GAN framework. The architecture of the proposed generator {$G$} is shown in Fig.\ref{fig:G}. It serves three functions: low-resolution spectral-spatial feature extraction, residual learning with contiguous memory mechanism, and super-resolution HSI reconstruction. \par  

Firstly, a feedforward 3D convolutional filter was introduced for the low-resolution spectral-spatial feature extraction. Unlike traditional RGB image super-resolution approaches that use 2D convolutional filters for spatial feature extraction, the HSI super-resolution requires processing continuous spectral channels and capturing spectral-spatial joint features from a data cube. Therefore, a 3D convolutional filter is a better choice for modelling both the spectral correlation characteristics and spatial texture information, and yielding the feature representation of the spectral-spatial hierarchical structures from low-resolution HSIs. In this study, the convolutional kernel is set to $16 \times 16 \times b$ for a $b$ band HSI input. Nah \textit{et al.} \cite{nah2017deep} found that the batch normalisation (BN) layer would normalise the features and get rid of the range flexibility for upscaling features. Therefore, no BN layer is used here to avoid blurring the spectral-spatial information hidden in the convolutional features. \par

Secondly, residual blocks (ResBlocks) were employed for the residual learning. The architecture of a ResBlock is shown in Fig.\ref{fig:G}b. Each ResBlock comprises two 3D convolutional filters, a ReLu activation layer, and a scaling layer. Wherein, two 3D convolutional filters are used to alleviate the effect of noise and strengthen the spectral-spatial hierarchical characteristics at the deeper level. The ReLU activator between convolutional filters is used to regularise deep features. The scaling layer after the last convolution layer in the ResBlock is used to scale the features, which increases the diversity of spectral-spatial hierarchical modes in the deep features. Similar to the initial feature extraction, no BN layer is used in the ResBlocks. The skip connection of ResBlocks forms an STSSRW mechanism, in which the spectral-spatial features extracted from previous ResBlocks are skip-connected with the features extracted from current ResBlocks. The skip-connected features not only represent extracted local spectral-spatial invariances, but also have the memory for high-frequency spectral-spatial information represented by a high dynamic range of previous features. This STSSRW mechanism enriches the spectral-spatial details with hierarchical structures and stabilises the learning process. It is noteworthy that for a given ResBlock, each convolutional layer with $m$ features and $n$ kernels has $O(m \times n \times 2)$ parameters, requiring $O(m\times n)$ memory. Although increasing the number of ResBlocks is an efficient way to improve the performance of a generator in extracting  deep features, it will lead to a high memory requirement. Moreover, the learning process may become numerically unstable with the increasing number of ResBlocks \cite{arjovsky2017towards}. Therefore, in this study, we set the number of ResBlocks to 34 to balance the model performance and cost.  \par 

Finally, the spectral-spatial feature maps extracted from multiple ResBlocks were fed into an UpscaleBlock to generate the super-resolution spectral-spatial features in the super-resolution HSI reconstruction. As shown in Fig.\ref{fig:G}c, the UpdscaleBlock is a combination of a 3D-convolutional filter and a shuffle layer, in which convolutional filters with a depth of 32 are used to exploit $k^2 \cdot 32 \cdot b$ features for an upscaling factor $k$. The shuffle layer is used to arrange all the features corresponding to each sub-pixel position in a pre-determined order and aggregate them into super-resolution areas. The size of each super-resolution area is $k \times k$. After this operation, the final feature maps with a size of $(k^2 \cdot 32 \cdot b) \times (H/k) \times (W/k)]$ will be arranged into the super-resolution feature maps with a size of $32 \cdot b \times H \times W$, where $H$ and $W$ are the height and width of the super-resolution HSI, respectively. At last, a deconvolution filter is used to decode the feature maps in each area, yielding the super-resolution HSI with enhanced spectral-spatial fidelity.

\subsubsection{The architecture of the discriminator $D$}
\label{sec:D} 
\begin{figure}[!t]   
    \centering  
    \includegraphics[width=3.6in]{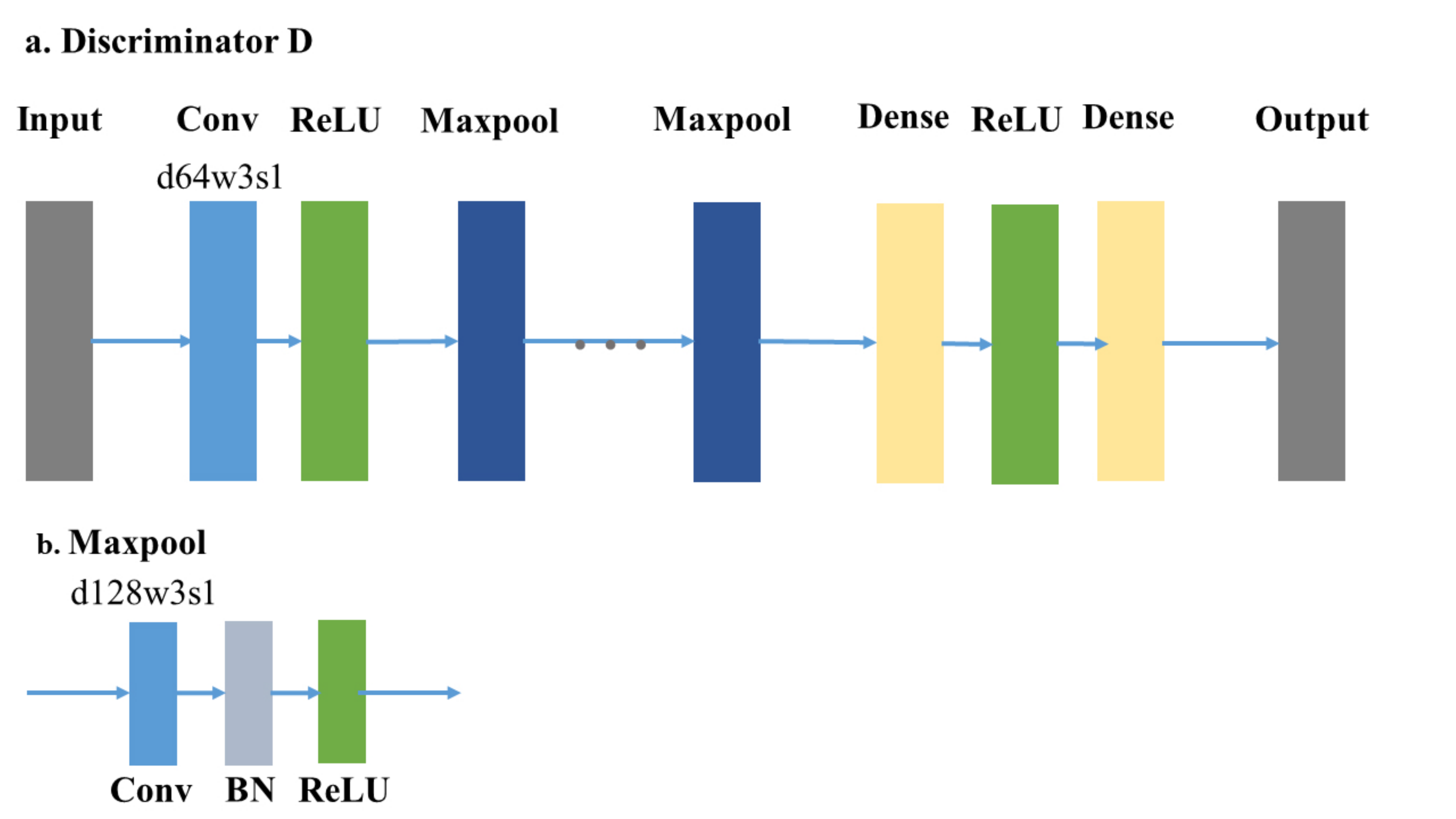}   
    \caption{The architectures of (a) the discriminator, D, and (b) the Maxpool block. Note: d,w and s denote the kernel depth, the kernel width and the stride of a convolutional layer, respectively.}    
    \label{fig:D}  
\end{figure}

The architecture of the proposed discriminator, $D$, as shown in Fig. \ref{fig:D}a, adopts an architecture similar to that used in \cite{RN2}. But, there is no sigmoid layer in our model, because the latent space optimisation requires the raw membership without compression. Thus, the proposed $D$ mainly contains one convolutional layer, $n$ Maxpool blocks ($n=8$ in this study) and two dense layers. The Maxpool block is a combination of a convolutional layer, a BN layer, and a ReLU layer (see Fig. \ref{fig:D}b). The Maxpool block aims to extract the high-level features of input data, and the resultant feature maps are input into two dense layers to obtain a membership distribution of the feature maps for real or generated HSIs. \par

\subsubsection{The architecture of the latent encoder, $L_E$}
The latent encoder, $L_E$, is developed and introduced to the GAN architecture for preventing mode collapse by mapping the generated spectral-spatial features from the image space to the latent space and produces the latent regularisation components in the learning process. Mathematically, we regard the spectral-spatial features as $SSF$, the singular value decomposition of $SSF$ in the latent space can be written as: 
\begin{equation}
	SSF = U \times \gamma \times V^T 
\label{eq:1-1}
\end{equation}

where, $U$ and $V$ are the left and right singular vectors of $SSF$, respectively, and $\gamma$ can be expressed as: 
\[\gamma=\begin{pmatrix}
SSD&0\\
0&0\\
\end{pmatrix}
\]

where $SSD = diag{\lambda_1, \lambda_2, ... , \lambda_r}$ represents the spectral-spatial distribution of $SSF$. If the mode collapse occurs in the learning process, the $SSD$ will concentrate on the first singular value $\lambda_1$ and the rest singular values would close to zero. Therefore, in order to avoid the mode collapse, we use a latent encoder to automatically generate a latent regularisation components for $SSF$ of the real data and the generated data, denoted as $L_E(I^{hr})$ and $L_E(G_{\theta_G}(I^{lr}))$, to compensate the singular value of $SSD$.

\begin{figure}[!t]   
    \centering  
    \includegraphics[width=3.6in]{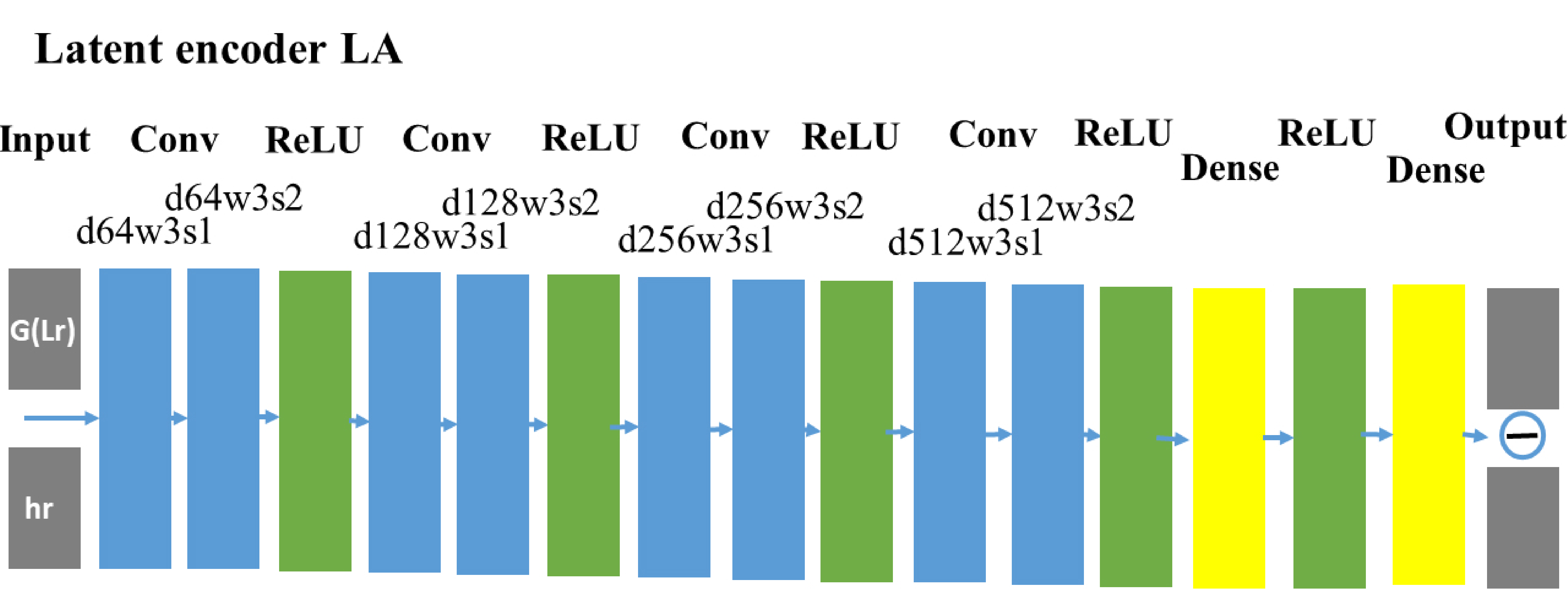}   
    \caption{The architecture of the latent encoder $L_E$. Note: d,w and s denote the kernel depth, the kernel width and the stride of a convolutional layer, respectively.}    
    \label{fig:La}  
\end{figure}

The architecture of the latent encoder is shown in Fig. \ref{fig:La}, which consists of eight convolutional layers with an increasing kernel depth by a factor 2 through different layers from 64 to 512. The striding operation is used to reduce the number of features once the kernel depth is doubled. The resultant of 512 feature maps are input into two dense layers so that its outputs match the dimension of the HSI. As shown in Fig. \ref{fig:Networkarchitecturte}, $La$ receives signals from the generator, $G(I^{lr})$, and the targeted data, $I^{hr}$. The outputs of the encoder are used to calculate an L2 loss to regularise the loss function of the discriminator, defined as:
\begin{equation}
	L_E = \mathbb{E}_{I^{hr} \sim Pr(I^{hr})}\|L_E(I^{hr})-L_E(G_{\theta_G}(I^{lr}))  \|_2
\label{eq:1}
\end{equation}

where $\left\lVert \cdot \right\rVert_2$ denotes L2 norm, $L_E(I^{hr})$ and $L_E(G_{\theta_G}(I^{lr}))$ are the output regularisation components in the latent space, respectively, corresponding to the inputs from real data and the generated data from the discriminator, $G_{\theta_G}$, parametrised by $\theta_G$. The encoder is simultaneously optimised with the generator, $G_{\theta_G}$. 
To make sure that the outputs of $L_E$ and the real high-resolution HSI in the latent space have the same dimension, $L_E$ is pre-trained by real HSI data. This speeds up the formal optimisation process.

\subsection{Model optimisation with spectral-spatial realistic perceptual loss}
\label{sec:Wloss}

In this study, we treat a low-resolution image as a low-dimension manifold embedded in the latent space, thus the super-resolution HSI can be generated by the parametrised latent space learnt by the model. Theoretically, the generated super-resolution sample, $G_{\theta_G}(I^{lr})$ , from a low-resolution sample, $I^{lr}$, by the generator will be located in a neighbourhood area of its target, $I^{hr}$, in the latent space.\par
Previous studies \cite{chen2018attention, zhang2020supervised, RN29} used the difference between $G_{\theta_G}(I^{lr})$ and $I^{hr}$ as the generator loss function, described as:
\begin{equation}
    \lvert G_{\theta_G}(I^{lr}) - I^{hr} \rvert_1 \leq \epsilon 
\label{eq:2_1}
\end{equation}

However, there are two drawbacks to use this loss function in the HSI super-resolution optimisation process. Firstly, the activated features in the latent space are very sparse. The distance based losses rarely consider the spectral-spatial consistency between $G_{\theta_G}(I^{lr})$ and $I^{hr}$, which leads to the spectral-spatial distortion in the generated super-resolution HSI results. Secondly, the direct bounding on the difference between $G_{\theta_G}(I^{lr})$ and $I^{hr}$ makes it hard to converge because $I^{lr}$ is usually disturbed by the network impairments or random noise.

In order to overcome the aforementioned drawbacks, we have designed a spectral-spatial realistic perceptual (SSRP) loss to comprehensively measure the spectral-spatial consistency between $G_{\theta_G}(I^{lr})$ and $I^{hr}$ in the latent space. The formula of the SSRP loss is defined as the weighted sum of the spectral contextual loss, the spatial texture loss, the adversarial loss, and a latent regularisation component, and is shown as follows: 
\begin{equation}
    L_G^{SSRP} = \lambda \cdot L_{spectral} + \eta \cdot L_{spatial} + \sigma \cdot L_{adversarial} + \mu \cdot L_E
\label{eq:SSRP}
\end{equation}
where $L_{spectral}$ is the spectral contextual loss, $L_{spatial}$ is the spatial texture loss, $L_{adversarial}$ is the adversarial loss, and $L_E$ is the latent regularisation component. \par

Based on the SSRP loss, the min-max problem in the GAN model can be described as follows:
\begin{equation}
\begin{aligned}
 \mathop{min}\limits_{\theta_G} \mathop{max}\limits_{\theta_D} L(D_{\theta_D}, G_{\theta_G }) 
 = \mathop{min}\limits_{\theta_G} \mathop{max}\limits_{\theta_D} (\lambda \cdot L_{spectral} \\ + \eta \cdot L_{spatial} + \sigma \cdot L_{adversarial} + \mu \cdot L_E)
\end{aligned}
\label{eq:5}
\end{equation}

The details of  $L_{spectral}$, $L_{spatial}$, $L_{adversarial}$, and $L_E$ are provided below.  \par

\subsubsection{Spectral contextual loss}

$L_{spectral}$ is designed to measure the spectral directional similarity between $G_{\theta_G}(I^{lr})$ and $I^{hr}$ in the latent space, which is defined as follows: 
\begin{equation}
    L_{spectral} = \mathbb{E}_{z}\{ -log(\frac{1}{N} \cdot \sum_j max_i A_{ij}) \}
\label{eq:Spectral}
\end{equation}

\begin{equation}
    A_{ij} = \frac{e^{1-b_{ij}/n_b}}{\sum_k e^{1-b_{ij}/n_b}}
\label{eq:Spectral-1}
\end{equation}

\begin{equation}
    b_{ij} = \frac{c_{ij}}{min c_{ik}}
\label{eq:Spectral-2}
\end{equation}

\begin{equation}
    c_{ij} = \frac{(D_{\mu}(G_{\theta_G}(I_{ij}^{lr}))-D_{\mu}(I_{ij}^{hr})) \cdot (I_{ij}^{lr}-I_{ij}^{hr})}{\Vert D_{\mu}(G_{\theta_G}(I_{ij}^{lr}))-D_{\mu}(I_{ij}^{hr}) \Vert_2 \cdot \Vert I_{ij}^{lr}-I_{ij}^{hr} \Vert_2}
\label{eq:Spectral-3}
\end{equation}

where $n_b$ is the band number of an HSI, $D_{\mu}(\cdot)$ denotes the feature maps obtained from the convolutional layer before the first Maxpooling layer of the discriminator, $D$. \par

\subsubsection{Spatial texture loss}

In GAN models, if the loss function only measures the spatial resemblance of the generated and targeted samples, it usually leads to the blurry super-resolution results. In this study, we introduce a spatial texture loss $L_{spatial}$ to measure the texture differences between the feature maps of $G_{\theta_G}(I^{lr})$ and $I^{hr}$ in the latent space. In the $L_{spatial}$, the feature maps of $G_{\theta_G}(I^{lr})$ and $I^{hr}$ before activation are used because they contain more sharp details. $L_{spatial}$ is defined as:  
\begin{equation}
    L_{spatial} = \mathbb{E}_{z}\{\frac{1}{W \cdot H} \cdot \sum_{i=1}^{W} \sum_{j=1}^{H} \Vert D_{\phi} (G_{\theta_G}(I_{ij}^{lr}))- D_{\phi} (I_{ij}^{hr}) \Vert_2 \}
\label{eq:spatial}
\end{equation} 
where $D_{\phi}(\cdot)$ denotes the  feature maps obtained from the convolutional layer after the last Maxpooling layer of the discriminator $D$.

\subsubsection{Adversarial loss}

Along with the spectral contextual loss and the spatial texture loss, an adversarial loss is introduced to facilitate the generator $G$ in reconstructing the image in the ambient manifold space, and fooling the discriminator network. $L_{adversarial}$ is defined based on the Wasserstein distance \cite{LN04} between the probability distributions of real data, $P_r(I^{hr})$, and the generated data, $P_g(G_{\theta_G}(I^{lr}))$. Theoretically, $L_{adversarial}$ is strong in alleviating the mode collapse during the training process, because the Wasserstein distance evaluating the similarity between $P_r(I^{hr})$ and $P_g(G_{\theta_G}(I^{lr}))$ rely on the whole samples distributions rather than the individual sample. In other words, there is a penalty would be triggered when the $P_g(G_{\theta_G}(I^{lr}))$ only covers a fraction of $P_r(I^{hr})$, which facilitates the diversity of the generated super-resolution HSI. The goal of $L_{adversarial}$ is to minimise the Wasserstein distance, $W_d(P_r,P_g)$, which is defined as:
\begin{equation}
\begin{aligned}
    L_{adversarial} = W_d(P_{r}, P_{g}) = \frac{1}{K} \mathop{sup} \limits_{{\Vert f \Vert}_L<K} \mathbb{E}_{I^{hr} \sim Pr(I^{hr})}[f(I^{hr})] \\ -\mathbb{E}_{I^{lr} \sim Pg(I^{lr})}[f(G_{\theta_G}(I^{lr})]
\end{aligned}
\label{eq:2}
\end{equation}
where $f$ is the K-Lipschitz function. Suppose we have a parametrised family of functions, ${\{f_{w_d}\}}_{w_d \in W_d}$, that are all K-Lipschitz for some $K$, then the $L_{adversarial}$ can be written as:
\begin{equation}
\begin{aligned}	
  L_D(P_r,P_g)=\mathop{max}\limits_{w\in W_d} \mathbb{E}_{I^{hr}\sim Pr(I^{hr})} [f_{w_d} (I^{hr})]\\
  -\mathbb{E}_{I^{lr}\sim Pg(I^{lr})} [f_{w_d} (G_{\theta_G}(I^{lr}))]
\end{aligned}
\label{eq:3}
\end{equation}
where $W_d$ is chosen such that the Lipschitz constant of $f_{w_d}$ is smaller than a constant, $K$. If the probability densities of $P_r (I^{hr})$ and $P_g (I^{lr})$ satisfy the Lipschitz continuous condition (LCC) \cite{RN34}, there is a solution $f_{w_d}$. Thus, the discriminator is trained to learn a K-Lipschitz continuous function to help compute the Wasserstein distance. The LCC is a strong pre-requisite for calculating $W_d(P_r,P_g)$. Therefore, the parameters, $w_d$, should lie in a $W_d$-dimensional manifold in order to meet this constraint. 

\subsubsection{The latent regularisation component}

In our proposed model, $G$ is a ResNet with global Lipschitz continuity. As described in Section \ref{sec:March}-c, we have introduced a latent encoder, $La$, to compensate the singular values of the spectral-spatial features of $I^{lr}$ to the desired $I^{hr}$. In addition to the optimisation process,  the Lipschitz Continuity Condition (LCC) is employed to enforce the  local spectral-spatial invariances of $G$, and map the latent manifold space to a more regularised latent space in case of mode collapse, described as:
\begin{equation}
    \lvert G_{\theta_G}(I^{lr}) - I^{hr} \rvert_1 \leq K \times \Vert La(G_{\theta_G}(I^{lr})) - La(I^{hr}) \Vert_2 
\label{eq:2_2}
\end{equation} 

Thus, the proposed loss function $L_E$ (see Eq. \ref{eq:1}) would be penalised if the singular values of the spectral-spatial features of a generated super-resolution HSI are updated in a particular direction. In other words, LLC-derived $L_E$ updating is able to prevent the learning process of each layer from becoming sensitive to the limited direction, which mathematically alleviates the mode collapse, in turn stabilising the optimisation process.  \par

\section{Experimental Evaluation}

In this section, we evaluate the effect of proposed LE-GEN and determined whether it will improve the super-resolution quality and facilitate manifold mapping for solving the problem of mode collapse. Wherein, the developed SSRP loss function plays a key role for both of these prospects. A total of three experiments are designed. The first experiment is to evaluate the optimal parameter combination for the SSRP loss in our proposed model, the second experiment is proposed to evaluate the super-resolution quality, and the last experiment is to evaluate the mode collapse in the model training. \par

The proposed model was trained and tested on real HSI datasets coming from different sensors. It was also compared with five state-of-the-art super-resolution models, including the hyperspectral coupled network (HyCoNet) \cite{zheng2020coupled}, the low tensor-train rank (LTTR) network \cite{dian2019learning}, the band attention GAN (BAGAN) \cite{RN14}, the super resolution generative adversarial network (SRGAN) \cite{RN2}, and the Wasserstein GAN (WGAN) \cite{LN04}. Among them, HyCoNet, LTTR and BAGAN are the state-of-the-art models for HSI super-resolution, while SRGAN and WGAN are the most widely used GAN frameworks for image super-resolution. In order to fit the HSI into the SRGAN and WGAN models, a band-wise strategy was employed \cite{mei2017hyperspectral}. 

\subsection{HSI data descriptions}
In our experiments, two types of datasets obtained from different sensors were used, one from the public AVIRIS archive, the other from the privately measured UHD-185 data of Guyuan Potato Field (GPF).
\subsubsection{AVIRIS datasets}
Two publicly available HSIs from the AVIRIS data archive were chosen, including the HSIs of Indian Pines (IP) data and the Kennedy Space Center (KSC). Each of them contains $224$ hyperspectral bands from $400 \sim 2500 nm$. The HSIs in the KSC dataset were collected by the Kennedy Space Center, Florida, on March 23, 1996. The spatial resolution was 18 m. The HSIs of IP covered the crop planting areas with the spatial resolution of 20 m in North-Western Indiana, USA. In this study, to keep the spectral consistency between different datasets, only the wavelength ranges from invisible to near-infrared ($450 \sim 950 nm$) were considered in our experiments.

\subsubsection{UHD-185 dataset}
The UHD-185 dataset contained three privately measured HSIs, denoted as $GPF-1$, $GPF-2$, and $GPF-3$, in Guyuan Potato Field, Hebei, China. Each of the HSIs was collected by the DJI S1000 UAV system (SZ DJI Technology Co Ltd., Gungdong, China) based UHD-185 Imaging spectrometer (Cubert GmbH, Ulm, Baden-Württemberg, Germany) in 2019. All the images were obtained at a flight height of 30 m, with 220 bands from visible to near-infrared bands between $450$ and $950$ nm and a spatial resolution close to 0.25m per pixel.  

\subsection{Evaluation metrics}
The evaluation metrics include 1) the metrics for evaluating super-resolution quality and robustness and 2) the metrics for evaluating mode collapse of GANs. 

\subsubsection{Evaluation metrics for super-resolution quality and robustness assessment}   

In total, five spectral-spatial evaluation metrics were employed for the super-resolution quality assessment.

 These five metrics are 1) Information entropy associated peak signal-to-noise (PSNR), 2) Spatial texture associated structural similarity index (SSIM), 3) Perception-distortion associated perceptual index (PI), 4) spectral reality associated spectral angle mapper (SAM) and 5) Spectral consistency associated spectral relative error (SRE). Among them, the PSNR and SSIM were widely used in the evaluation of image quality \cite{RN41}, the larger the score of PSNR or SSIM the higher the image-quality. 
 
 The PSNR is defined as:
\begin{equation}
PSNR(I^{hr},I^{sr})=10 \cdot \log_{10}(255^2/MSE(I^{hr},I^{sr}))
\end{equation}
where $MSE(I^{hr},I^{sr})$ is the mean squared error between the real HR HSI, $I^{hr}$, and the generated HR HSI through super-resolution, $I^{sr}$. The PSNR goes to infinity as the MSE goes to zero. \par

The SSIM is defined as:
\begin{equation}
SSIM(I^{hr},I^{sr})=l(I^{hr},I^{sr}) \cdot c(I^{hr},I^{sr}) \cdot s(I^{hr},I^{sr})
\end{equation}
where $l(I^{hr},I^{sr})$, $c(I^{hr},I^{sr})$, and $s(I^{hr},I^{sr})$ are the difference measures for luminance, contrast, and saturation between real and generated HR HSI pairs, respectively. The details can be found in \cite{RN44}. \par

However, the numerical scores of PSNR and SSIM are not always correlated well with the subjective image quality. Therefore, Blau \textit{et al.} \cite{RN45} proposed an index, PI (Perception Index), as a compensatory reference for the image quality evaluation. The lower the PI value is, the higher the perceptual quality of the image. The PI is defined by two non-referenced image quality measurements, MA \cite{RN46} and NIQE \cite{RN47}, described as:
\begin{equation}
PI(I^{hr},I^{sr})=\frac{1}{2}((10-MA(I^{hr},I^{sr}))) + NIQE(I^{hr},I^{sr})
\end{equation}

In order to measure the spectral distortion, the spectral angle mapper(SAM), was used to calculate the average angle between a super-resolution HSI and its targeted high-resolution HSI. The SAM is defined as: 
\begin{equation}
SAM(I^{hr},I^{sr})=\frac{1}{n} \sum arccos(\frac{I^{hr} \cdot I^{sr}}{\Vert I^{hr} \Vert_2 \cdot \Vert I^{sr} \Vert_2})
\end{equation}
where $n$ is the pixel number of the HSI.\par

To evaluate the pixel-wised spectral reconstruction quality, the spectral relative error (SRE) was also used as a metric, defined as: 
\begin{equation}
SRE(I^{hr},I^{sr})=[\frac{1}{n_b} \sum^{n_b}_{i=1} \Vert (I_{i}^{hr},I_{i}^{sr}) \Vert^2]^{\frac{1}{2}}
\end{equation}

where the $n_b$ is the band number of an HSI.

\subsubsection{Evaluation metrics for mode collapse of GANs }
Two metrics for GANs, Inception Score (IS) and Frechet Inception Distance (FID), were employed to measure the mode collapse through monitoring the image quality and diversity in the model training process \cite{liu2019spectral,hartmann2018eeg}. The IS measures both the image quality of generated HSIs and their diversity, reflecting the probability of mode collapse in the model training process. In GANs, it is desirable for the conditional probability, $p(I^{hr}|G(I^{lr}))$ to be highly predictable (low entropy), that is, the probability density function is less uniform. The diversity of the generated image can be measured with the marginal probability, $p(I^{hr})= \int p(I^{hr}|G(I^{lr}))d{I^{lr}}$. The less uniform (low entropy) the marginal probability is, the less the diversity of the generated image is. Through computing the KL-divergence between these two probability distributions, the IS is computed with the equation below:
\begin{equation}
    IS=exp[\mathbb{E}_{I^{lr}\sim p(I^{lr})}[\mathbb{D}_{KL}(p(I^{hr}|G(I^{lr}))||p(I^{hr}))]]
\end{equation}

The Frechet Inception Distance (FID) score is a metric calculating the distance between the feature vectors extracted from real and generated images.  It was used to evaluate the quality of GAN generated images, and a lower score correlates with a higher image quality. The lower the FID value is, the better image quality and diversity are. The FID is sensitive to mode collapse.  Through modelling the distribution of the features extracted from an intermediate layer with a multivariate Gaussian distribution, the FID between the real image and generated images is calculated using the following equation,
\begin{equation}
	FID=||M_{hr} - M_{sr}||_2^2 + Tr(C_{hr} + C_{sr} - 2(C_{hr} \times C_{sr})^{1/2})
\end{equation}

where $M_{hr}$ and $M_{sr}$ refer to the feature-wise means of the real high-resolution HSI and the generated super-resolution HSI in discriminator model, respectively, and $C_{hr}$ and $C_{sr}$ are the covariance matrix for the real and generated feature vectors, respectively.

\subsection{Experimental configuration}\label{experimentconfiguration}

In our experiments, the raw HSIs were labelled as HR samples. The LR samples were generated by down-sampling the HR samples with three scaling factors, $\times 2$, $\times 4$ and $\times 8$, based on the bi-cubic interpolation approach \cite{RN38}. For the AVIRIS datasets, the KSC data was used for the model training and test, and the IP data was used for the independent test. For the UHD-185 dataset, the $GPF-1$ and $GPF-2$ were used for training and test, and the $GPF-3$ was used for the independent test. More specifically, for the training/test datasets, the HR HSI was cropped into a series of sub-images with a spatial size of $384 \times 384$, and the corresponding LR data was respectively cropped to $192 \times 192$, $96 \times 96$, and $48 \times 48$. After this operation, a total of 196 HR and LR HSI pairs were generated from the AVIRIS dataset, and 352 HR and LR HSI pairs were generated from the UHD-185 dataset, in which $70\%$ of image pairs were randomly selected as the training set and the rest $30\%$ of image pairs were used as the test set. \par

The training process was divided into two stages. In the first stage, the discriminator $D$ and the latent encoder $La$ were pre-trained over 5,000 iterations on the raw HR HSI dataset to get initial weights. The Adam optimiser was used by setting the forgetting factors, $\beta_1=0.9$ and $\beta_2=0.999$, a small scalar constant $\epsilon=10^{-7}$ and the learning rate $= 10^{-4}$ \cite{RN39}. In the second stage, the discriminator, the generator, and the latent encoder were jointly trained for over 10,000 times, until they converged. The Adam optimiser with the same parameters was used. All of the training were performed on NVIDIA 1080Ti GPUs. We also investigated the effect of the hyper-parameter $\lambda$ on the optimisation performance. We found a $\lambda$ value in the range of $0 \sim 0.1$ could generate high-quality HSI data. In this study, $\lambda = 0.01$ was used for training the proposed model. \par

\subsection{Experiment 1: the parameter selection for the SSRP loss function}

To achieve an optimal performance, an optimised combination of the parameters in the SSRP loss function Eq.\eqref{eq:SSRP}, $\lambda$, $\eta$, $\sigma$ and $\mu$, needs to be found. In this study, a traversal method was employed to search the optimal parameter combination. These parameters were traversed in the range of $0$ to $100$ with a fixed step of $0.001$ for the range of $0$ to $1$, and a fixed step of $0.1$ for the range of $1$ to $100$. The selection of parameter combinations was based on the spectral-spatial quality of generated super-resolution HSIs measured with five evaluation metrics, PSNR, SSIM, PI, SAM, and SRE. Table. \ref{table:paremeter} lists the top five parameter combinations and the corresponding values of these metrics for generating the super-resolution HSI with the scaling factors of $\times 2$, $\times 4$ and $\times 8$. It can be observed that all the parameters after optimisation are located in a relatively small range, for example, $12.3-12.8$ for $\lambda$ and $\eta$, $0.004-0.009$ for $\sigma$ and $0.014-0.017$ for $\mu$. In the following experiments, we employed the average values of the best parameters for various scaling factors, thus, $\lambda = 12.5, \eta = 12.5, \sigma = 0.0063$, and $ \mu = 0.015$.

\begin{table}[]
\caption{The top five combinations of the parameters, $\lambda$, $\eta$, $\sigma$ and $\mu$ for the super-resolution HSI generation with the scaling factors of $\times 2$, $\times 4$ and $\times 8$ based on the values of spectral-spatial evaluation metrics after 10,000 iterations.}
\label{table:paremeter} 
\centering
\resizebox{3.6in}{!}{
\begin{tabular}{cccccccc}
\toprule
Scaling factor & No. & ($\lambda$, $\eta$, $\sigma$, and $\mu$) & PSNR            & SSIM           & PI             & SAM            & SRE          \\ \midrule
                 & 1   & (12.8,  12.9,  0.008,  0.015)          & \textbf{31.738} & \textbf{0.982} & \textbf{3.782} & \textbf{5.011} & \textbf{8.383} \\
                 & 2   & (12.8,  12.8,  0.009,  0.016)          & 31.716          & 0.945          & 3.884          & 5.155          & 8.461          \\
$\times 2$       & 3   & (12.7,  12.8,  0.007,  0.014)          & 31.712          & 0.963          & 3.87           & 5.115          & 8.469          \\
                 & 4   & (12.8,  12.8,  0.008,  0.014)          & 31.712          & 0.943          & 3.849          & 5.161          & 8.482          \\
                 & 5   & (12.6,  12.8,  0.006,  0.017)          & 31.708          & 0.926          & 3.876          & 5.174          & 8.499          \\ \hline
                 & 1   & (12.4,  12.4,  0.006,  0.015)          & \textbf{31.417} & \textbf{0.903} & 3.765          & \textbf{4.942} & \textbf{8.219} \\
                 & 2   & (12.4,  12.5,  0.009,  0.014)          & 31.395          & 0.901          & \textbf{3.764} & 5.075          & 8.267          \\
$\times 4$       & 3   & (12.4,  12.3,  0.007,  0.015)          & 31.375          & 0.893          & 3.765          & 5.013          & 8.279          \\
                 & 4   & (12.5,  12.8,  0.007,  0.014)          & 31.359          & 0.891          & 3.767          & 5.017          & 8.276          \\
                 & 5   & (12.5,  12.8,  0.006,  0.017)          & 31.322          & 0.898          & 3.819          & 5.065          & 8.331          \\ \hline
                 & 1   & (12.3,  12.3,  0.005,  0.015)          & \textbf{29.881} & \textbf{0.931} & 3.672          & \textbf{4.741} & \textbf{8.672} \\
                 & 2   & (12.4,  12.3,  0.006,  0.014)          & 29.851          & 0.902          & 3.663          & 4.828          & 8.726          \\
$\times 8$       & 3   & (12.4,  12.2,  0.004,  0.014)          & 29.816          & 0.885          & \textbf{3.583} & 4.797          & 8.753          \\
                 & 4   & (12.5,  12.5,  0.005,  0.014)          & 29.828          & 0.923          & 3.617          & 4.866          & 8.679          \\
                 & 5   & (12.4,  12.6,  0.005,  0.015)          & 29.791          & 0.885          & 3.634          & 4.817          & 8.733    \\ \bottomrule                       
\end{tabular}
}
\end{table}

\subsection{Experiment 2: model robustness and super-resolution quality assessment}
\label{exp:2}

To evaluate the robustness and generalizability of the proposed model, we have evaluated our model on both testing datasets and independent datasets. 

\subsubsection{Model assessment on the testing datasets} \label{sss:1}

As described in Section \ref{experimentconfiguration}, we divided the dataset into the training and testing datasets. The performance of the proposed model for hyperspectral super-resolution with three upscaling factors ($\times 2$, $\times 4$ and $\times 8$) was evaluated on testing datasets including AVIRIS (KSC) and UHD-185(GPF-1 and GPF-2), and compared with five state-of-the-art competition models. To assess the model robustness to noise, the models was also evaluated on the datasets artificially added with three Gaussian white noise levels ($\infty$, $40 db$ and $80 db$) to each of the spectral bands of low-resolution HSIs. To facilitate ranking the models in terms of reconstruction quality, five most widely used evaluation metrics, PSNR, SSIM, PI, SAM, and SRE, were chosen. Specifically, PSNR, SSIM and PI were used to measure the spatial reconstructed quality from the aspects of information entropy, spatial similarity, and perception distortion, respectively. The higher PSNR and SSIM scores and the lower PI scores indicate the higher spatial reconstruction quality. In addition, the SAM and SRE scores were used for the spectral distortion measurement from the aspects of spectral angle offset and amplitude difference, respectively. The lower values of SAM and SRE scores indicate the higher spectral reconstruction quality. \par

Table \ref{table:comp_AV} and Table \ref{table:comp_UHD} provide the average scores of PSNR, SSIM, PI, SAM, and SRE of HSI super-resolution results from the proposed model and its five competitors using the AVIRIS and UHD-185 testing datasets, respectively. 
In general, the results on both datasets consistently show that the proposed LE-GAN model achieves the highest PSNR and SSIM values and the lowest PI, SAM and SRE values for all three different upscaling factors and three added noise levels (see the highlighted values in Table \ref{table:comp_AV} and Table \ref{table:comp_UHD}). This means that LE-GAN achieves the best spectral and spatial fidelity and super-resolution quality.\par

A more detailed analysis of the results for the model performance evaluation was performed from two aspects: (1) Super-resolution performance under various upscaling factors (2) Model robustness against different noise levels. Since the results in Table \ref{table:comp_AV} and Table \ref{table:comp_UHD} have the similar patterns for all the models, here we only present the analyses and assessment using the results on AVIRIS data (i.e. Table \ref{table:comp_AV}):  \par 
(1) Among three upscaling factors, the LE-GAN based super-resolution with the smallest upscaling factor $\times 2$ and without added noise (i.e. $\infty$ db) achieves the best spectral and spatial reconstruction quality. The best scores of PSNR, SSIM, PI, SAM, and SRE are $35.513$, $0.898$, $3.052$, $4.207$, and $8.379$, respectively, which are closer to the real high-resolution HSI (i.e. $35.981$ for PSNR, $0.912$ for SSIM, $3.011$ for PI, $4.142$ for SAM, and $8.019$ for SRE), compared to its competitors. The similarities (i.e. the ratio between the super-resolution HSI and the real high-resolution HSI) reach $98.87\%$, $98.46\%$, $98.66\%$, $98.45\%$, and $95.7\%$, respectively.
In addition, for a given added noise level, the spectral and spatial quality of the LE-GAN generated super-resolution HSIs are more stable between the upscaling factors of $\times 2$ and $\times 4$. For example, under the added noise level of $80 db$, the PSNR, SSIM, PI, SAM and SRE scores are $35.225$, $0.835$, $3.171$, $4.221$, and $8.839$ for $\times 2$ upscaling factor, and increasing the upscaling factor to $\times 4$ only causes the slight changes to these scores which are $34.975$, $0.804$, $3.364$, $4.362$ and $9.062$, respectively. The consistency ratios (i.e. the ratio between the $\times 2$ and $\times 4$ super-resolution HSI) are $99.29\%$, $96.29\%$, $94.26\%$, $96.78\%$, and $97.54\%$, respectively. 
In contrast, a larger performance degradation occurs on the spectral and spatial reconstruction quality of the competitors. For example, with regard to the WGAN, the second best model in terms of PSNR and SSIM, the scores of PSNR, SSIM, PI, SAM, and SRE under non-added noise level are $33.729$, $0.826$, $3.867$, $7.248$, $14.152$ with $\times 2$ upscaling, but change to $30.035$, $0.807$, $4.476$, $7.922$, $14.361$ with $\times 4$ upscaling, showing the performance degradations of $10.95\%$, $2.3\%$, $13.6\%$, $8.5\%$, and $1.5\%$, respectively.  
Although the degradations of PSNR, SSIM, PI, SAM, SRE scores can be observed on all the models for $\times 8$ upscaling, the degradation rate of these scores from the proposed LE-GAN is the smallest. 
For example, under the non-added noise ($\infty$ db), the SNR, SSIM, PI, SAM, SRE scores of LE-GAN based super-resolution HSIs for $\times 8$ upscaling are $32.078$, $0.784$, $3.988$, $4.711$ and $8.986$, respectively, which are $21.03\%$, $6.76\%$, $31.4\%$, $15.9\%$ and $25.4\%$ higher than those based on the second best models (i.e. the WGAN in terms of SSIM ($26.591$) and the BAGAN in terms of PSNR ($26.591$), PI ($5.814$), SAM ($5.602$), and SRE($12.048$)).  \par

(2) With regard to the model robustness to noise, the proposed LE-GAN shows the best performance on the spectral and spatial reconstruction for a given upscaling factor in comparison with its competitors, although the degradation is observed with increased noise levels. 
The smaller the upscaling factor is, the more robust the model is. The most robust results against noise are at the upscaling factor of  $\times 2$. Only $3.6\%$, $7.9\%$, $14.67\%$, $13.11\%$. and $7.33\%$ degradations of the PSNR, SSIM, PI, SAM, and SRE scores of LE-GAN-based super-resolution results occur when the added noise level increases from non-added ($\infty$ db) to $40$ db (see Table \ref{table:comp_AV}). In contrast, the added noise-induced degradations to the results from the WGAN (the second best model for $\times 2$ upscaling factor) are much higher, reaching $20.29\%$, $8.35\%$, $26.65\%$, $20.23\%$, and $21.61\%$, respectively.
In addition, when the upscaling factor increases from $\times 2$ to $\times 8$, the added noise-induced degradations on the PSNR, SSIM, PI, SAM and SRE scores of the LE-GAN super-resolution results are $9.58\%$, $11.86\%$, $19.42\%$, $14.02\%$, and $11.25\%$, which are acceptable for the super-resolution with a high upscaling factor and high noises. In contrast, a more serious deterioration can be observed in the results from its competitors. For example, the added noise-induced degradations on the PSNR, SSIM, PI, SAM, SRE of the BAGAN-based super-resolution results, the second best model, are $21.72\%$, $10.57\%$, $15.6\%$, $13.45\%$ and $15.05\%$, respectively, for an upscaling factor of $\times 2$, but change to $24.07\%$, $23.12\%$, $13.23\%$, $15.55\%$, $19.84\%$ for an upscaling factor of $\times 8$ . \par

\begin{table}[]
\caption{A quantitative comparison of HSIs super-resolution spectral and spatial quality in terms of the average PSNR, SSIM, PI, SAM, SRE scores using the proposed model and five competition models on test datasets, AVIRIS (KSC Data), with various upscaling factors and added noise levels. Note that the PSNR/SSIM/PI/SAM/SRE scores for the KSC data (i.e. real high-resolution HSI) are $35.981$, $0.912$, $3.011$, $4.142$, and $8.019$ respectively. The higher PSNR, SSIM and the lower PI, SAM, SRE, the better the spectral and spatial fidelity.}
\label{table:comp_AV} 
\centering
\resizebox{3.6in}{!}{
\begin{tabular}{ccccccccc}
\toprule
                    & SNR(db)  &      & HyCoNet  & LTTR  & BAGAN  & SRGAN  & WGAN   & LE-GAN \\ \midrule
                    & $\infty$ & PSNR & 31.213 & 29.495 & 32.177 & 32.421 & 33.729 & \textbf{35.513}  \\
                    &          & SSIM & 0.792  & 0.729  & 0.766  & 0.809  & 0.826  & \textbf{0.898}   \\
                    &          & PI   & 4.181  & 4.269  & 3.672  & 4.015  & 3.867  & \textbf{3.052}   \\
                    &          & SAM  & 6.491  & 6.515  & 5.485  & 9.014  & 7.248  & \textbf{4.207}   \\
                    &          & SRE  & 10.813 & 10.145 & 10.476 & 15.438 & 14.152 & \textbf{8.379}   \\
                    & 80       & PSNR & 28.751 & 24.981 & 29.121 & 30.106 & 32.945 & \textbf{35.225}  \\
                    &          & SSIM & 0.719  & 0.735  & 0.761  & 0.756  & 0.808  & \textbf{0.835}   \\
AVIRIS ($\times 2$) &          & PI   & 4.178  & 4.819  & 3.766  & 3.991  & 4.196  & \textbf{3.171}   \\
                    &          & SAM  & 6.622  & 7.164  & 5.961  & 10.961 & 7.297  & \textbf{4.221}   \\
                    &          & SRE  & 12.447 & 11.913 & 11.216 & 19.301 & 17.914 & \textbf{8.839}   \\
                    & 40       & PSNR & 23.777 & 23.253 & 25.187 & 26.181 & 26.886 & \textbf{34.223}  \\
                    &          & SSIM & 0.571  & 0.634  & 0.685  & 0.715  & 0.757  & \textbf{0.827}   \\
                    &          & PI   & 4.751  & 3.182  & 4.351  & 5.489  & 5.272  & \textbf{3.577}   \\
                    &          & SAM  & 7.591  & 7.516  & 6.338  & 11.177 & 9.087  & \textbf{4.842}   \\
                    &          & SRE  & 14.574 & 12.946 & 12.332 & 20.028 & 18.053 & \textbf{9.042}   \\ \hline
                    & $\infty$ & PSNR & 27.216 & 27.841 & 29.177 & 26.105 & 30.035 & \textbf{35.367}  \\
                    &          & SSIM & 0.774  & 0.725  & 0.762  & 0.799  & 0.807  & \textbf{0.835}   \\
                    &          & PI   & 4.781  & 4.474  & 4.291  & 4.812  & 4.476  & \textbf{3.061}   \\
                    &          & SAM  & 6.514 & 6.533  & 5.566  & 9.729  & 7.922  & \textbf{4.272}   \\
                    &          & SRE  & 12.75 & 10.811 & 10.641 & 16.68  & 14.361 & \textbf{8.431}   \\
                    & 80       & PSNR & 24.816  & 25.896 & 27.048 & 22.777 & 28.183 & \textbf{34.975}  \\
                    &          & SSIM & 0.559  & 0.665  & 0.513  & 0.524  & 0.709  & \textbf{0.804}   \\
AVIRIS ($\times 4$) &          & PI   & 4.514  & 5.889  & 4.356  & 5.037  & 5.101  & \textbf{3.364}   \\
                    &          & SAM  & 6.571  & 7.031  & 5.685  & 11.582 & 8.926  & \textbf{4.362}   \\
                    &          & SRE  & 13.551 & 11.562 & 11.406 & 17.799 & 18.321 & \textbf{9.062}   \\
                    & 40       & PSNR & 21.714 & 22.669 & 26.363 & 25.934 & 29.093 & \textbf{33.041}  \\
                    &          & SSIM & 0.315  & 0.651  & 0.479  & 0.668  & 0.604  & \textbf{0.786}   \\
                    &          & PI   & 6.051  & 6.072  & 4.854  & 6.223  & 5.912  & \textbf{3.376}   \\
                    &          & SAM  & 6.771  & 7.736  & 6.268  & 12.091 & 10.138 & \textbf{4.443}   \\
                    &          & SRE  & 15.041 & 13.203 & 13.114 & 18.158 & 18.556 & \textbf{9.383}   \\ \hline
                    & $\infty$ & PSNR & 19.871 & 21.747 & 26.591 & 24.589 & 25.332 & \textbf{32.078}  \\              
                    &          & SSIM & 0.622  & 0.638  & 0.718  & 0.668  & 0.731  & \textbf{0.784}   \\
                    &          & PI   & 6.835  & 6.322  & 5.814  & 6.536  & 5.902  & \textbf{3.988}   \\
                    &          & SAM  & 7.69  & 6.992  & 5.602  & 10.179 & 8.126  & \textbf{4.711}   \\
                    &          & SRE  & 14.361 & 14.302 & 12.048 & 16.943 & 16.092 & \textbf{8.986}   \\
                    & 80       & PSNR & 17.821 & 20.402 & 24.073 & 20.603 & 24.369 & \textbf{30.291}  \\
                    &          & SSIM & 0.552  & 0.493  & 0.692  & 0.605  & 0.663  & \textbf{0.775}   \\
AVIRIS ($\times 8$) &          & PI   & 7.421  & 7.094  & 6.411  & 6.942  & 6.116  & \textbf{4.326}   \\
                    &          & SAM  & 7.72  & 7.667  & 6.283  & 11.499 & 9.321  & \textbf{4.732}   \\
                    &          & SRE  & 16.361 & 15.534 & 13.194 & 18.213 & 18.101 & \textbf{9.768}   \\
                    & 40       & PSNR & 14.84 & 18.344 & 20.191 & 19.532 & 21.572 & \textbf{29.003}  \\
                    &          & SSIM & 0.316  & 0.418  & 0.552  & 0.511  & 0.547  & \textbf{0.691}   \\
                    &          & PI   & 7.622  & 7.954  & 6.701  & 7.924  & 7.366  & \textbf{4.949}   \\
                    &          & SAM  & 9.172  & 9.061  & 6.616  & 12.541 & 8.525  & \textbf{5.479}   \\
                    &          & SRE  & 18.219 & 18.836 & 15.031 & 19.597 & 19.598 & \textbf{10.125}  \\ \bottomrule
\end{tabular}
}
\end{table}

\begin{table}[]
\caption{A quantitative comparison of HSIs super-resolution spectral and spatial quality in terms of the average PSNR, SSIM, PI, SAM, SRE scores using the proposed model and five competition models on test datasets, UHD-185 (GPF-1 and GPF-2), with various upscaling factors and added noises. Note that the PSNR/SSIM/PI/SAM/SRE scores for the KSC data (i.e. real high-resolution HSI) are $38.915$, $0.992$, $4.418$, $6.942$, and $10.519$ respectively. The higher PSNR and SSIM and the lower PI, SAM, and SRE, the better the spectral and spatial fidelity.}
\label{table:comp_UHD} 
\centering
\resizebox{3.6in}{!}{
\begin{tabular}{ccccccccc}
\toprule
                    & SNR(db)  &      & HyCoNet  & LTTR  & BAGAN  & SRGAN  & WGAN   & LE-GAN \\ \midrule
                    & $\infty$ & PSNR & 33.238 & 32.689 & 34.642 & 36.009 & 37.697 & \textbf{38.575}  \\
                    &          & SSIM & 0.874  & 0.875  & 0.851   & 0.897  & 0.879  & \textbf{0.979}   \\
                    &          & PI   & 5.11  & 5.454  & 4.799  & 5.185  & 5.207  & \textbf{4.323}   \\
                    &          & SAM  & 9.174  & 8.124  & 7.266  & 11.848 & 9.904  & \textbf{6.893}   \\
                    &          & SRE  & 15.755 & 14.711 & 12.677 & 17.341 & 15.43  & \textbf{10.295}  \\
                    & 80       & PSNR & 30.484 & 29.196 & 31.913 & 32.28  & 35.37  & \textbf{37.625}  \\
                    &          & SSIM & 0.796  & 0.831  & 0.841  & 0.846  & 0.873  & \textbf{0.922}   \\
UHD-185 ($\times 2$) &          & PI   & 5.911 & 5.113  & 5.002  & 5.341  & 5.523  & \textbf{4.238}   \\
                    &          & SAM  & 9.331  & 9.073  & 7.921  & 13.05  & 9.849  & \textbf{6.899}   \\
                    &          & SRE  & 14.355 & 13.441 & 13.039 & 20.549 & 19.842 & \textbf{10.711}  \\
                    & 40       & PSNR & 28.834 & 27.137 & 28.904 & 29.296 & 33.752 & \textbf{36.976}  \\
                    &          & SSIM & 0.593  & 0.632  & 0.758  & 0.776  & 0.833  & \textbf{0.873}   \\
                    &          & PI   & 5.905  & 6.223  & 5.636  & 6.536  & 6.574  & \textbf{4.592}   \\
                    &          & SAM  & 10.53  & 10.311 & 10.172 & 13.597 & 11.328 & \textbf{6.508}   \\
                    &          & SRE  & 16.029 & 14.603 & 14.523 & 21.425 & 19.695 & \textbf{10.823}  \\ \hline
                    & $\infty$ & PSNR & 29.855 & 30.341 & 34.111 & 33.285 & 34.863 & \textbf{38.303}  \\
                    &          & SSIM & 0.717  & 0.804  & 0.827  & 0.861  & 0.834  & \textbf{0.892}   \\
                    &          & PI   & 6.161  & 5.755  & 5.309  & 6.194  & 5.944  & \textbf{4.391}   \\
                    &          & SAM  & 9.408  & 9.187  & 7.311  & 12.652 & 9.937  & \textbf{6.563}   \\
                    &          & SRE  & 13.81  & 12.442 & 12.208 & 18.622 & 15.776 & \textbf{10.005}  \\
                    & 80       & PSNR & 27.046 & 27.702 & 30.396 & 30.731 & 33.866 & \textbf{37.061}   \\
                    &          & SSIM & 0.661  & 0.717  & 0.666  & 0.813  & 0.814  & \textbf{0.816}   \\
UHD-185 ($\times 4$)  &          & PI   & 6.791   & 6.995  & 5.741  & 6.061  & 6.276  & \textbf{4.463}   \\
                    &          & SAM  & 9.376  & 9.909  & 7.956  & 13.939 & 10.09  & \textbf{6.554}   \\
                    &          & SRE  & 15.464 & 14.963 & 17.128 & 21.509 & 19.706 & \textbf{10.361}   \\
                    & 40       & PSNR & 26.948 & 26.891 & 27.917 & 25.844 & 32.217 & \textbf{36.556}  \\
                    &          & SSIM & 0.459  & 0.606  & 0.567  & 0.751   & 0.687  & \textbf{0.731}   \\
                    &          & PI   & 7.521   & 7.197  & 6.136  & 7.621  & 7.311    & \textbf{4.751}   \\
                    &          & SAM  & 12.679 & 10.262 & 9.028  & 13.99  & 12.912 & \textbf{6.661}   \\
                    &          & SRE  & 16.518 & 14.396 & 14.385 & 22.221 & 20.411   & \textbf{10.939}  \\ \hline
                    & $\infty$ & PSNR & 22.675 & 25.375 & 27.831 & 27.367 & 29.012 & \textbf{35.397}  \\
                    &          & SSIM & 0.705  & 0.688  & 0.784  & 0.734  & 0.821  & \textbf{0.852}   \\
                    &          & PI   & 7.908  & 7.557  & 6.835  & 7.863  & 7.153  & \textbf{5.046}   \\
                    &          & SAM  & 9.487  & 9.668  & 7.827  & 13.942 & 10.251  & \textbf{7.104}   \\
                    &          & SRE  & 16.378 & 16.287 & 13.64  & 19.011 & 17.271  & \textbf{10.145}  \\
                    & 80       & PSNR & 21.879 & 23.88  & 26.586 & 24.029 & 27.604 & \textbf{33.166}  \\
                    &          & SSIM & 0.677  & 0.549  & 0.564  & 0.679  & 0.755  & \textbf{0.801}   \\
UHD-185 ($\times 8$) &          & PI   & 8.541   & 8.258  & 7.426  & 8.393  & 7.611   & \textbf{5.649}  \\
                    &          & SAM  & 9.814  & 10.52  & 8.931   & 14.092 & 12.152 & \textbf{7.669}   \\
                    &          & SRE  & 17.627 & 16.941 & 15.946 & 22.224 & 21.887 & \textbf{11.333}  \\
                    & 40       & PSNR & 17.819 & 21.166 & 23.199 & 22.251  & 24.132 & \textbf{31.519}  \\
                    &          & SSIM & 0.419  & 0.474  & 0.541  & 0.591   & 0.632  & \textbf{0.715}   \\
                    &          & PI   & 8.674  & 9.154  & 7.872  & 9.058  & 8.755  & \textbf{6.229}   \\
                    &          & SAM  & 13.694 & 11.175 & 8.559  & 15.236 & 14.212 & \textbf{7.945}   \\
                    &          & SRE  & 20.042 & 20.604 & 16.103 & 25.159 & 22.086 & \textbf{11.162}  \\ \bottomrule
\end{tabular}
}
\end{table}

\subsubsection{Model assessment on the independent test datasets}

\begin{figure}[]   
	\centering  
	\includegraphics[width=3.6in]{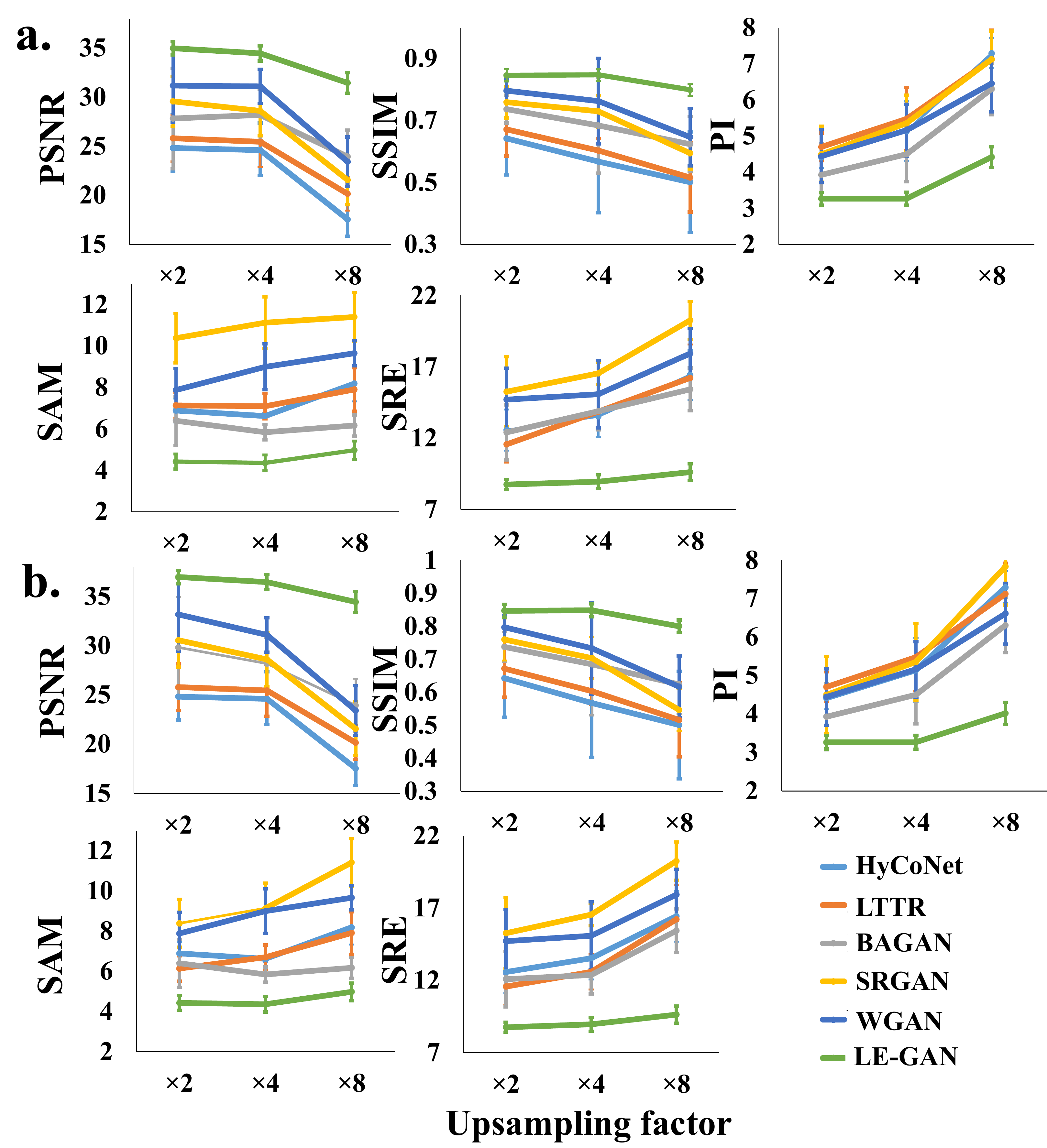}   
	\caption{A comparison of five evaluation metrics (PSNR, SSIM, PI, SAM, and SRE) between the proposed model and its five competitors evaluated on independent test datasets (a) AVIRIS (IP) and (b) UHD-186 (GPF-3). The average values and standard deviations of each metrics are calculated across three different noise levels($\infty$, $80 db$ and $40 db$). }    
	\label{fig:state_analysis_1}  
\end{figure}

The proposed model has also been evaluated on two independent test datasets, AVIRIS (IP) and UHD-185 (GPF-3), which were not involved in the model training. Fig. \ref{fig:state_analysis_1} illustrates a comparison of five evaluation metrics (PSNR, SSIM, PI, SAM and SRE) between the proposed model and its five competitors. The average value and standard deviation of each metric were calculated based on the measures at three noise levels, $\infty$, $80 db$ and $40 db$. Compared to its competitors, the proposed model achieves the highest average values and lowest standard deviations for PSNR, SSIM, and the lowest average values and the lowest standard deviation for PI, SAM and SRE, across three upscaling factors on both AVIRIS test dataset (see Fig. \ref{fig:state_analysis_1}a) and UHD-185 test dataset (see Fig. \ref{fig:state_analysis_1}b). That is, the proposed model achieves the best performance on super-resolution. Similar to the evaluation results in Subsection \ref{sss:1}, overall the second best model on the independent test datasets is WGAN for the spatial information reconstruction measure (e.g. PSNR, SSIM), and BAGAN for the spectral information reconstruction measure (e.g. SAM, SRE). 

It can also be observed that the changes of these metrics are relatively small with the increase of the upscaling factor. When the upscaling factor increases from $\times 2$ to $\times 4$, the average values of SSIM, PI, SAM, and SRE from the proposed model almost stay the same; When the upsampling factor increases from $\times 4$ to $\times 8$, the changes of these metrics are much smaller compared to those from its competitors.  

These findings suggest that the proposed model overcomes the drawback associated with spectral-spatial reconstruction under the noises interferences compared to its competitors. Moreover, the proposed model is less sensitive to the upscaling factor, and has a good performance even with a large upscaling factor (e.g. $\times 8$).   \par

\subsubsection{Visual Analysis of generated super-resolution HSIs with a large upsampling factor ($\times 8$)}

\begin{figure}[]   
\centering  
\includegraphics[width=3.6in]{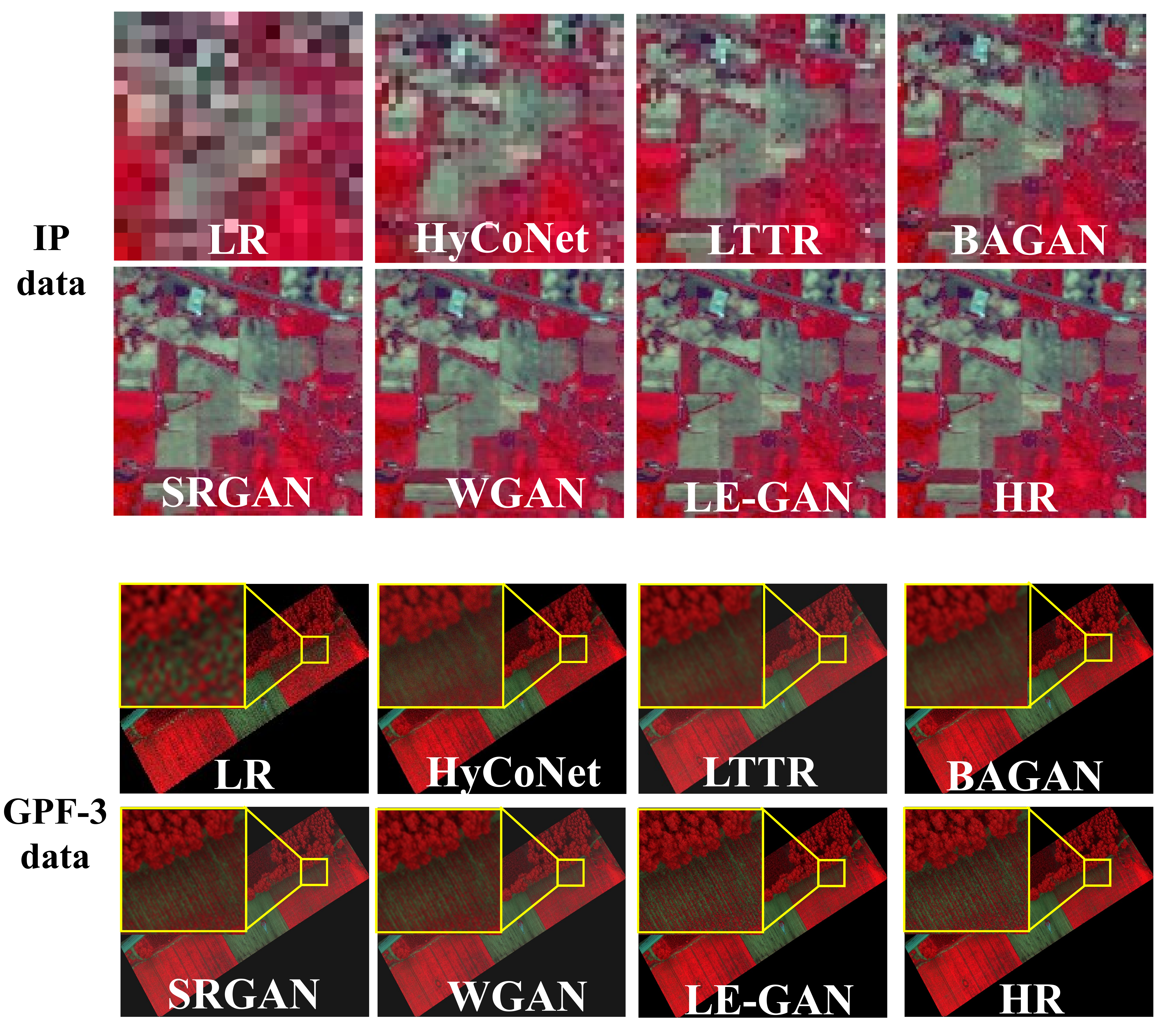}   
\caption{A sample of super-resolution results ($\times 8$) from our models and its five competitors on independent test datasets}    
\label{fig:EVA_img}  
\end{figure}

To demonstrate the performance improvement of the proposed model in spectral-spatial fidelity, visual analyses on generated super-resolution HSI samples have been performed. Fig. \ref{fig:EVA_img} displays the results from independent test datasets (IP and GPF-3).
Although the visualisation results from the proposed method and its competitors are similar, the image edges from the LE-GAN are sharper than those from the competitors. For example, the internal textures of the bare-soil shown as grey in the false-colour images almost disappear in the generated super-resolution IP images by the HyCoNet, LTTR, and BAGAN (the second, third, and forth images in the first row of Fig. \ref{fig:EVA_img}). These findings suggest that the LE-GAN provides improved spatial quality in general. 


\begin{figure}[]   
    \centering  
    \includegraphics[width=3.6in]{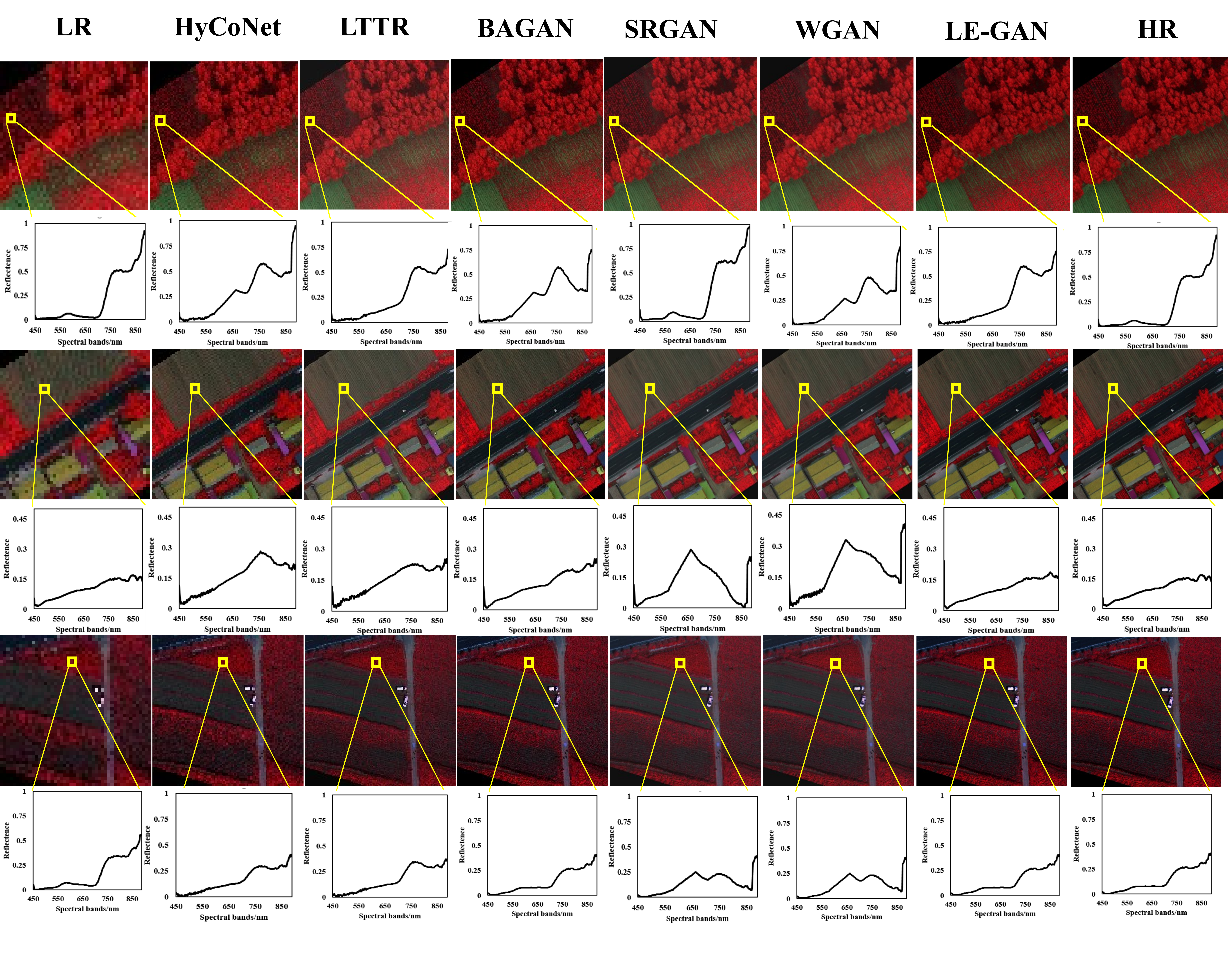}   
    \caption{Detailed spectral analysis on local patches of false colour super resolution ($\times 8$) results generated by different models from the independent test dataset, GPF-3. The results in each column from the left to right are for real low resolution (LR) HSI patches, high resolution images generated from models (SRCNN, SRResNet, VDSR, SRGAN, WGAN, and the proposed LRE-GAN), and the corresponding High resolution (HR) HSI patches, respectively.}  
    \label{fig:patch_img}  
\end{figure}

\begin{figure}[]   
    \centering  
    \includegraphics[width=3.6in]{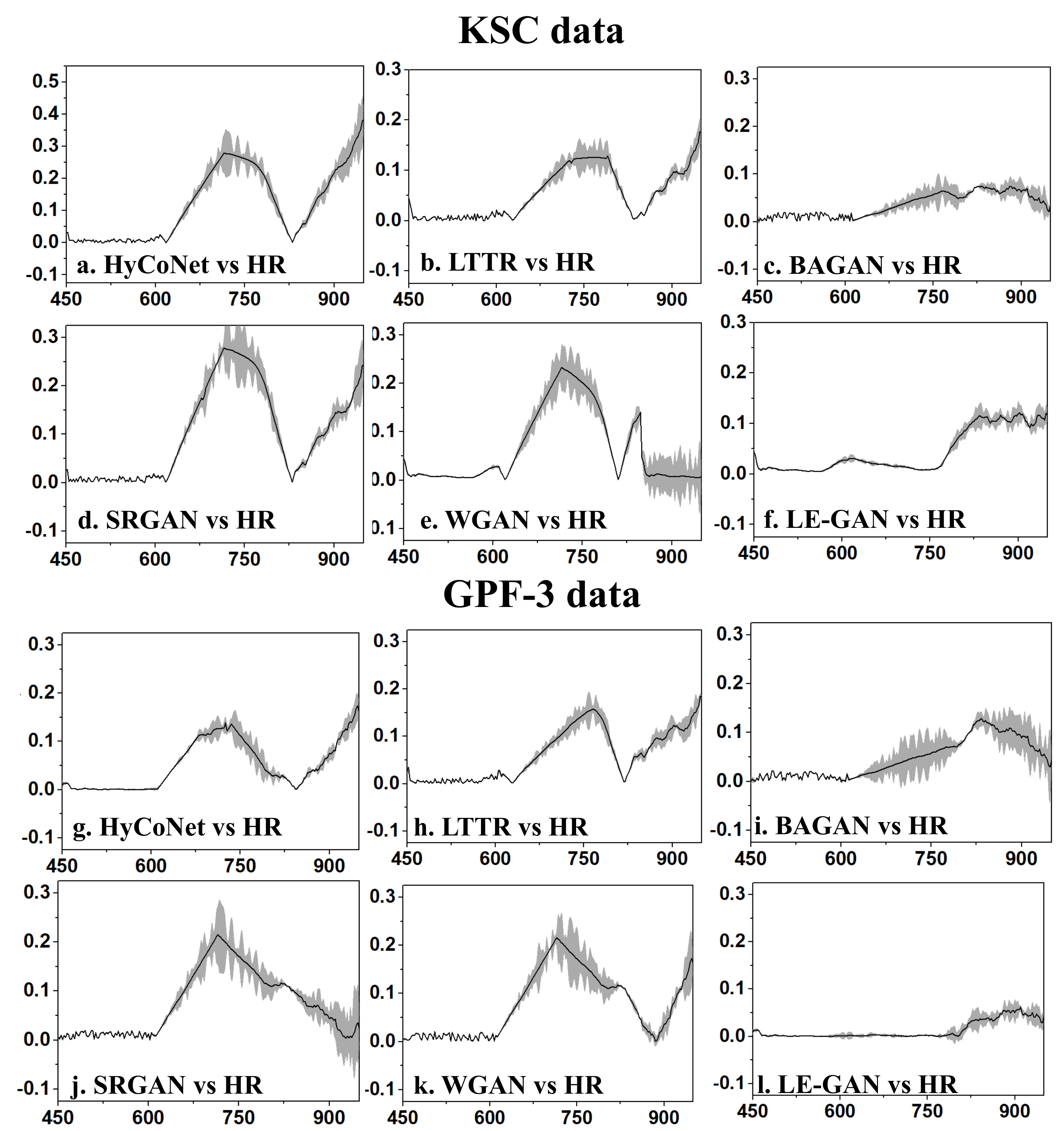}   
    \caption{A spectral residual (the black line) and deviation (the grey shadow) analysis between real HR HSI and the generated super-resolution HSI from different models on the model test dataset (KSC data) and the independent test dataset (GPF-3 data)}    
    \label{fig:spec_consis}  
\end{figure}

To further visualise the super-resolution details on spatial and spectral fidelity, some representative false-colour composite image patches and the spectral curves of the super-resolved HSI patches from independent test dataset (GPF-3) are shown in Fig. \ref{fig:patch_img}. It is obvious that the brightness, contrast, and internal structures of the false-colour images generated by the LE-GAN are more faithful to real HR data. For example, the land cover textures in the LE-GAN generated image (the second image from the right in the second row of images in Fig. \ref{fig:patch_img}) are clearer, compared to the images generated by the competitors (e.g. the HyCoNet and LTTR based images) in which the edges of streets are fuzzy. Moreover, the spectral curves from the LE-GAN generated images are more consistent with those from real HR HSI data. For example, the typical vegetation spectral curves in the images generated by the HyCoNet, SRGAN, and WGAN reveal distinct biases in the range of red-edge to near-infrared with real HR data (see the images of the first row in Fig. \ref{fig:patch_img}). In contrast, the vegetation spectral curves from the LE-GAN super-resolution are more consistent with those from real HR HSI. A detailed analysis of the spectral residual and standard deviation between the generated HSI and real HR HSI from the independent test dataset is shown in Fig. \ref{fig:spec_consis}. It can be found that the residual error between the LE-GAN generated HSI and HR HSI is close to zero in the range of 450 to 780 nm and lower than $0.25$ in the range of 780 to 950 nm, and the deviation is lower than $0.02$. All these results suggest that the proposed model provides a better performance in HSI super-resolution without losing the spectral details. The second and third best spectral residuals are achieved by the BAGAN and LTTR, respectively, and the spectral biases in the range of 630 to 950 nm and the average deviations reach $0.029$ and $0.041$, respectively. 

\subsection{Experiment 3: Mode collapse evaluation}

Generally, mode collapse mainly happens in the training process when the super-resolution HSI produced by the generator only partially covers the target domain. When the generator learns a type of spectral-spatial pattern that is able to fool the discriminator, it may keep generating this kind of pattern so that the learning process is over-learned. The distance and distribution of the generated super-resolution HSI provide the most direct evidence for determining whether the mode collapse occurs in the generator. In this section, we evaluated the effect of the proposed LE-GAN on alleviating mode collapse from three aspects: 1) a quantitative evaluation on the diversity of the generated super-resolution HSI, based on the distance-derived IS and FID metrics, 2) a smoothness monitoring on the generator iterations during the network training process, and 3) a visualisation of the distributions of the real high-resolution HSI samples and the generated super-resolution samples.  \par

Firstly, the quantitative evaluation for the diversity of the generated super-resolution HSI was conducted on the testing dataset and independent dataset mentioned in Section \ref{experimentconfiguration}. In addition, in order to assess the potential affects of different upscaling factors and added-noise levels on the occurrence of mode collapse, all of the experiments were conducted on three upscaling factors ($\times 2$, $\times 4$ and $\times 8$) with three Gaussian white noise levels ($\infty$, $40 db$ and $80 db$), and compared with five state-of-the-art competition models. The IS and FID were used as the evaluation metrics for assessing the diversity of the super-resolution HSIs and determining the existence of a mode collapse. A higher IS and lower FID scores will show the better diversity of the generated super-resolution HSI and the sign of the alleviation of mode collapse.  \par

Table \ref{table:mode_collapse} lists the IS and FID measurements on the proposed LE-GAN and five selected competitors using the testing datasets. The evaluation results on AVIRIS and UHD-185 testing datasets demonstrate that the proposed LE-GAN model outperforms its competitors in terms of the IS and FID measurements for all three upscaling factors and added noise combinations (see the highlighted values in Table \ref{table:mode_collapse}). They also indicate that the proposed model has greater performance on alleviating mode collapse issue occurred in the generated spectral-spatial diversity. The differences of these measurements between the proposed model and its competitors are particularly significant for the cases of low added noise and low upscaling levels. For example, the proposed LE-GAN achieves the highest IS ($13.46$ for AVIRIS data and $14.69$ for UHD-185 data) and the lowest FID ($13.7$ for AVIRIS data and $37.95$ for UHD-185 data) on the datasets with the $\times 2$ upscaling factor and non-added noise of $\infty$ SNR. \par

In addition, Table \ref{table:mode_collapse} also reveals the IS and FID degradations with the increase of upscaling factor and added noise level for all of the models. The comparison of these degradations can help to explore the model robustness in preventing the mode collapse issue. Specifically, an average of $33.6\%$ IS drop and $32.03\%$ FID increase are observed from the LE-GAN-based super-resolution results when increasing the upscalling factor from $\times 2$ to $\times 8$ and decreasing SNR from from $\infty$ to $40 db$. Meanwhile, the results from the WGAN, the second best model in terms of super-resolution fidelity (see Section \ref{exp:2}), show an average of $59.08\%$ IS drop and $40.97\%$ FID increase.
These findings suggest that the proposed LE-GAN achieves the best performance in preventing mode collapse under the higher upscaling factor and noise interferences.\par

\begin{table}[]
\caption{A Comparison of Inception Scores (IS) and Frechet Inception Distances (FID) of super-resolution HSIs generated from the proposed model and five competition models using the model test datasets.}
\label{table:mode_collapse} 
\centering
\resizebox{3in}{!}{
\begin{tabular}{ccccccccc}
\toprule
&      &        & \multicolumn{2}{l}{AVIRIS} & \multicolumn{2}{l}{UHD-185} \\
Upscaling & SNR      &        & IS               & FID         & IS           & FID          \\ \hline
            &          & HyCoNet  & 11.63            & 57.37       & 9.64         & 80.15        \\
            &          & LTTR  & 11               & 45.55       & 8.61         & 78.67        \\
            & $\infty$ & BAGAN  & 12.26            & 48.81       & 12.01        & 70.34        \\
            &          & SRGAN  & 10.62            & 53.88       & 11.06        & 74.05        \\
            &          & WGAN   & 13.25            & 24.37       & 13.66        & 49.13        \\
            &          & LE-GAN & \textbf{13.46}   & \textbf{13.7}   &\textbf{14.69}   & \textbf{37.95}        \\
            &          & HyCoNet  & 7.63         & 58.59       & 6.07         & 96.89        \\
            &          & LTTR  & 6.79         & 50.23       & 7.99         & 77.26        \\
2         & 80       & BAGAN  & 7.95         & 49.87       & 7.21         & 83.95        \\
            &          & SRGAN  & 6.77         & 54.59       & 6.09         & 92.06        \\
            &          & WGAN   & 11.37        & 24.45       & 8.12         & 60.83        \\
            &          & LE-GAN & \textbf{11.56}   & \textbf{15.86}  & \textbf{12.16}    & \textbf{40.34}        \\
            &          & HyCoNet  & 4.35         & 63.01       & 4.05         & 104.3        \\
            &          & LTTR  & 5.02         & 52.42       & 4.9          & 80.12        \\
            & 40       & BAGAN  & 6.32         & 49.59       & 4.79         & 87.1         \\
            &          & SRGAN  & 5.04         & 60.55       & 4.61         & 94.97        \\
            &          & WGAN   & 9.17         & 26.71       & 5.67         & 76.48        \\
            &          & LE-GAN & \textbf{10.16}   & \textbf{18.94}  & \textbf{10.13}    & \textbf{49.53}        \\ \hline
            &          & HyCoNet  & 9.79         & 62.14       & 8.87         & 87.65        \\
            &          & LTTR  & 9.82         & 49.46       & 7.1          & 86.22        \\
            & $\infty$ & BAGAN  & 11.03        & 53.51       & 10.69        & 77.25        \\
            &          & SRGAN  & 9.64         & 58.82       & 12.78        & 80.93        \\
            &          & WGAN   & 11.14        & 26.25       & 11.87        & 75.45        \\
            &          & LE-GAN & \textbf{12.22}   & \textbf{15.37}  & \textbf{14.41}   & \textbf{40.76}        \\
            &          & HyCoNet  & 6.19         & 63.88       & 5.35         & 105.44       \\
            &          & LTTR  & 5.91         & 54.81       & 7.08         & 84.31        \\
4         & 80       & BAGAN  & 6.49         & 53.74       & 6.52         & 91.22        \\
            &          & SRGAN  & 5.4          & 58.94       & 5.42         & 101.1        \\
            &          & WGAN   & 9.89         & 26.71       & 7.35         & 77.32        \\
            &          & LE-GAN & \textbf{11.28}   & \textbf{17.84}  & \textbf{12.77}    & \textbf{43.86}        \\
            &          & HyCoNet  & 3.83         & 68.34       & 2.91         & 113.66       \\
            &          & LTTR  & 4.32         & 57.13       & 4.49         & 87           \\
            & 40       & BAGAN  & 5.07         & 54.2        & 4.35         & 94.9         \\
            &          & SRGAN  & 4.39         & 66.08       & 3.72         & 103.29       \\
            &          & WGAN   & 7.55         & 28.64       & 5.22         & 83.75        \\
            &          & LE-GAN & \textbf{10.43}    & \textbf{19.91}  & \textbf{10.31}    & \textbf{50.53}        \\ \hline
            &          & HyCoNet  & 8.44         & 64.71       & 8.04         & 91.34        \\
            &          & LTTR  & 8.97         & 52.06       & 6.53         & 89.81        \\
            & $\infty$ & BAGAN  & 9.33         & 55.47       & 9.31         & 81.3         \\
            &          & SRGAN  & 8.85         & 61.81       & 10.92        & 84.82        \\
            &          & WGAN   & 9.95         & 26.7        & 10.93        & 79.14        \\
            &          & LE-GAN & \textbf{11.67}   & \textbf{16.42}  & \textbf{12.11}   & \textbf{42.4}         \\
            &          & HyCoNet  & 4.94         & 66.62       & 4.35         & 110.48       \\
            &          & LTTR  & 4.57         & 57.24       & 5.71         & 88.31        \\
8         & 80       & BAGAN  & 5.96         & 55.75       & 5.3          & 95.53        \\
            &          & SRGAN  & 4.54         & 61.99       & 4.1          & 106.1        \\
            &          & WGAN   & 8.89         & 28          & 6.1          & 81.33        \\
            &          & LE-GAN & \textbf{10.21}    & \textbf{19.76}  & \textbf{11.84}    & \textbf{45.92}        \\
            &          & HyCoNet  & 3.02         & 71.35       & 2.25         & 119.21       \\
            &          & LTTR  & 3.92         & 59.84       & 3.91         & 91.52        \\
            & 40       & BAGAN  & 4.32         & 56.21       & 3.12         & 98.9         \\
            &          & SRGAN  & 3.18         & 69.28       & 2.69         & 108.21       \\
            &          & WGAN   & 6.47         & 39.15       & 4.51         & 88.02        \\
            &          & LE-GAN & \textbf{9.6}     & \textbf{20.31}  & \textbf{9.03}    & \textbf{55.42}        \\ \bottomrule
\end{tabular}
}
\end{table}

Table \ref{table:mode_collapse_independent} provides the scores of the IS and FID from the proposed LE-GAN and its five competitors using the independent datasets. Similar to the results shown in Table \ref{table:mode_collapse}, the results highlighted in Table \ref{table:mode_collapse_independent} illustrate that the LE-GAN model achieves the best and most robust performance in terms of IS and FID for all the upscaling factor and added noise combinations. In the case of a $\times 2$ upscaling factor and non-added noise, the LE-GAN achieves the best IS and FID measurements (IS of $12.91$ and FID of $14.95$ for AVIRIS and IS of $15.27$ and FID of $39.22$ for UHD-185 dataset), with the smallest IS drop ($39.1\%$) and FID increase ($31.55\%$). These results are consistent with the mode collapse assessment reported in Table \ref{table:mode_collapse}, suggesting that the LE-GAN derived super-resolution HSIs have the best spectral-spatial diversity with alleviated mode collapse. \par

\begin{table}[]
\caption{A Comparisons of Inception Scores (IS) and Frechet Inception Distances (FID) from the proposed model and five competition models using the independent test datasets.}
\label{table:mode_collapse_independent} 
\centering
\resizebox{3in}{!}{
\begin{tabular}{ccccccccc}
\toprule
&      &        & \multicolumn{2}{l}{AVIRIS} & \multicolumn{2}{l}{UHD-185} \\ \cline{4-7} 
Upscaling & SNR  &        & IS           & FID         & IS           & FID          \\ \hline
            &      & HyCoNet  & 11.31        & 59.85       & 11.72        & 81.51        \\
            &      & LTTR  & 11.58        & 48.61       & 9.71         & 80.69        \\
            & $\infty$ & BAGAN  & 11.28        & 52.14       & 12.58        & 72.74        \\
            &      & SRGAN  & 10.91        & 57.35       & 13.25        & 76.32        \\
            &      & WGAN   & 12.22        & 26.26       & 14.45        & 70.71        \\
            &      & LE-GAN & \textbf{12.91} & \textbf{14.95} & \textbf{15.27} & \textbf{39.22}        \\
            &      & HyCoNet  & 7.12         & 61.86       & 6.93         & 98.2         \\
            &      & LTTR  & 6.97         & 53.77       & 9.09         & 79.18        \\
2         & 80   & BAGAN  & 8.32         & 52.03       & 8.48         & 85.1         \\
            &      & SRGAN  & 6.72         & 58.12       & 6.7          & 94.95        \\
            &      & WGAN   & 9.89         & 26.77       & 9.31         & 73.04        \\
            &      & LE-GAN & \textbf{10.91} & \textbf{16.51} & \textbf{13.47}  & \textbf{41.66}        \\
            &      & HyCoNet  & 4.07         & 65.47       & 5.1          & 107.41       \\
            &      & LTTR  & 5.34         & 55.72       & 7.38         & 82.46        \\
            & 40   & BAGAN  & 5.34         & 52.91       & 5.56         & 88.16        \\
            &      & SRGAN  & 5.13         & 63.8        & 4.76         & 97.11        \\
            &      & WGAN   & 7.64         & 27.33       & 7.02         & 79.78        \\
            &      & LE-GAN & \textbf{9.18} & \textbf{19.79}  & \textbf{11.4}  & \textbf{50.11}        \\ \hline
            &      & HyCoNet  & 9.1          & 65.06       & 10.97        & 89.56        \\
            &      & LTTR  & 9.62         & 52.16       & 8.53         & 88.38        \\
            & $\infty$& BAGAN  & 9.89         & 56.53       & 11.57        & 79.97        \\
            &      & SRGAN  & 9.46         & 62.35       & 13.65        & 83.64        \\
            &      & WGAN   & 10.41        & 27.73       & 14.56        & 77.62        \\
            &      & LE-GAN & \textbf{10.76}   & \textbf{14.8}  & \textbf{14.18}    & \textbf{42.2}         \\
            &      & HyCoNet  & 5.49         & 67.05       & 6.27         & 107.44       \\
            &      & LTTR  & 6.07         & 58.44       & 7.45         & 86.43        \\
4         & 80   & BAGAN  & 6.68         & 56.39       & 7.26         & 93.15        \\
            &      & SRGAN  & 5.09         & 62.95       & 6.04         & 103.94       \\
            &      & WGAN   & 8.74         & 28.27       & 8.3          & 79.41        \\
            &      & LE-GAN & \textbf{9.94}  & \textbf{18.35}    &\textbf{11.84}  & \textbf{45.02}        \\
            &      & HyCoNet  & 3.85         & 71.67       & 4.26         & 116.89       \\
            &      & LTTR  & 4.7          & 60.31       & 6.37         & 89.89        \\
            & 40   & BAGAN  & 5.61         & 57.67       & 3.85         & 96.17        \\
            &      & SRGAN  & 4.3          & 69.84       & 4.11         & 105.25       \\
            &      & WGAN   & 6.57         & 29.18       & 6.86         & 86.78        \\
            &      & LE-GAN & \textbf{8.3}   & \textbf{21.05}  & \textbf{10.99}  & \textbf{54.84}        \\ \hline
            &      & HyCoNet  & 8.4          & 68.3        & 9.06         & 93.81        \\
            &      & LTTR  & 8.22         & 55.14       & 7.37         & 92.68        \\
            & $\infty$ & BAGAN  & 9.27         & 58.74       & 10.25        & 83.57        \\
            &      & SRGAN  & 8.31         & 65.12       & 11.83        & 87.25        \\
            &      & WGAN   & 9.31         & 28.2        & 12.29        & 81.61        \\
            &      & LE-GAN & \textbf{9.8}   & \textbf{15.84}   & \textbf{12.82}    & \textbf{43.43}        \\
            &      & HyCoNet  & 4.53         & 70          & 5.47         & 113.15       \\
            &      & LTTR  & 4.61         & 60.32       & 6.73         & 90.34        \\
8         & 80   & BAGAN  & 5.58         & 58.78       & 5.78         & 97.65        \\
            &      & SRGAN  & 4.51         & 65.9        & 5.13         & 108.75       \\
            &      & WGAN   & 8.81         & 29.62       & 6.85         & 83.75        \\
            &      & LE-GAN & \textbf{9.17}  & \textbf{18.05}  & \textbf{10.16}       & \textbf{47.84}        \\
            &      & HyCoNet  & 3.37         & 75.23       & 3.33         & 122.07       \\
            &      & LTTR  & 4.44         & 63.42       & 4.23         & 94.2         \\
            & 40   & BAGAN  & 3.98         & 59.71       & 3.96         & 101.72       \\
            &      & SRGAN  & 3.03         & 73.26       & 3.54         & 110.76       \\
            &      & WGAN   & 6.7          & 31.34       & 5.76         & 90.11        \\
            &      & LE-GAN & \textbf{8.77}   & \textbf{22.03}  & \textbf{8.2}     & \textbf{56.81}        \\ \bottomrule
\end{tabular}
}
\end{table}
    
\begin{figure}[]   
	\centering  
	\includegraphics[width=3.6in]{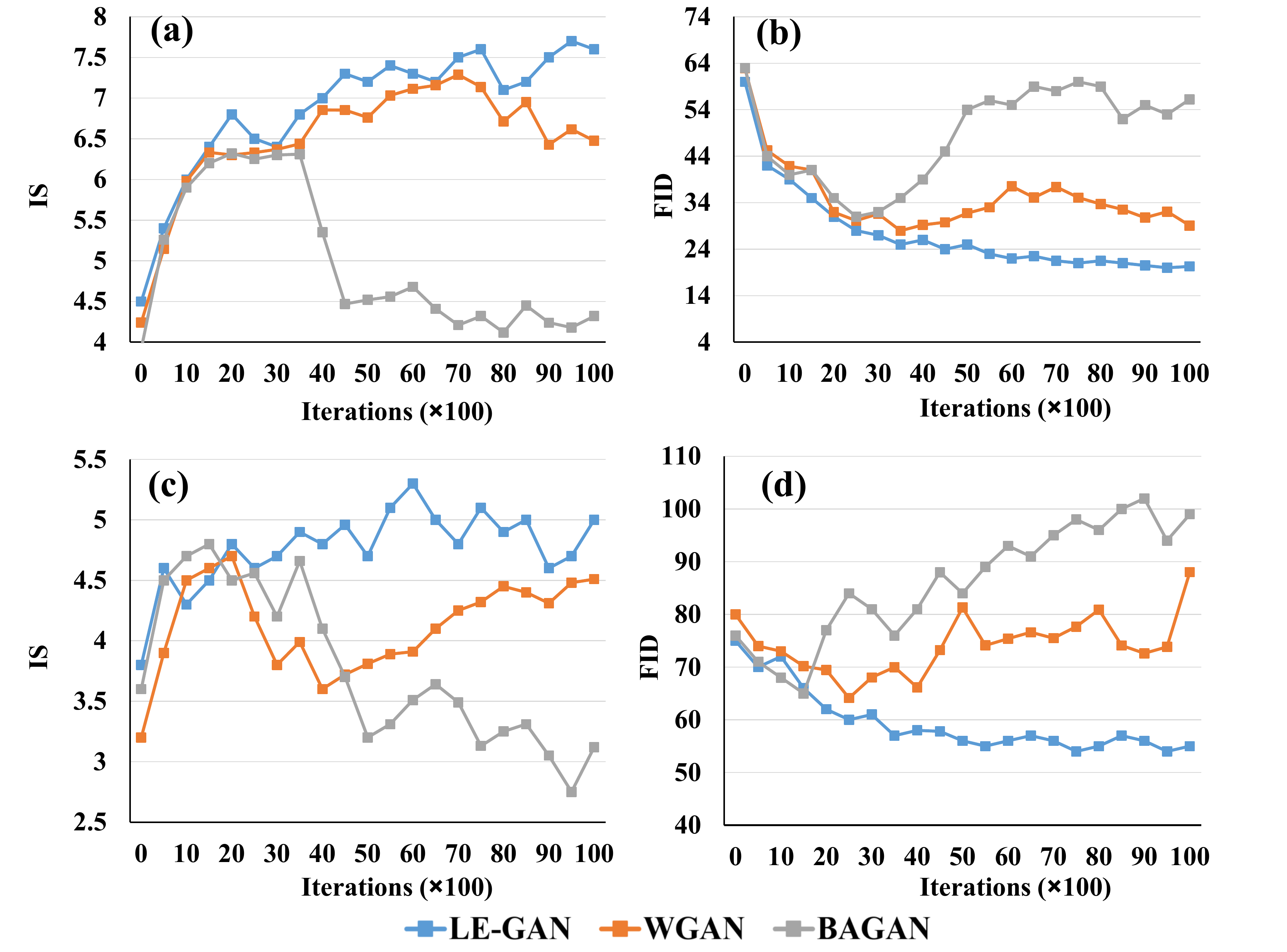}   
	\caption{The changes of IS and FID scores during the training of the proposed LE-GAN and its two best competitors (WGAN and BAGAN). The model training is conducted on (a-b) AVIRIS and (c-d) UHD-186 training dataset with a SNR level of 40db and an upscaling factor of $\times 8$. }    
	\label{fig:mode_collapse_1}  
\end{figure}

Secondly, a smoothness monitoring on the generator iteration was used to determine if mode collapse occurred during the training process. According to the results illustrated in Table \ref{table:mode_collapse} and Table \ref{table:mode_collapse_independent}, the high noise-levels and the large upscaling factors lead to more serious mode collapse. To demonstrate the performance difference between the proposed model and its competitors in alleviating model collapse, a comparison was conducted  under a high added noise level (e.g. $40db$) and a high upscaling factor (e.g. $\times 8$). Fig. \ref{fig:mode_collapse_1} illustrates the IS and FID iterations of the generated HSIs from the proposed LE-GAN and the two best competitors, i.e. WGAN and BAGAN, during the training process. It is obvious that the IS and FID curves from the proposed model are smoother and more stable than those from WGAN or BGAN, along with the increase of iteration number. Unlike the curves from WGAN or BGAN, the curves of IS from the proposed mode, LE-GAN, steadily increase and the curves of FID steadily decrease for both AVIRIS and UHD-186 datasets. This indicates that there is no significant mode collapse occurs during the training of LE-GAN. However, a big drop of IS is observed during the training of BAGAN (e.g. after 3500 iterations, as shown in Fig. \ref{fig:mode_collapse_1}a ) and during the training of WGAN (e.g. after 2000 iterations, as shown in Fig. \ref{fig:mode_collapse_1}c). Moreover, the curves of FID don't steadily decrease during the training for WAGAN or BAGAN. These observations indicate that the mode collapse occurs in the training of representative GAN models (e.g. WGAN and BAGNAN), and the proposed model is more effective in alleviating the mode collapse. \par

Finally, to further understand and assess the performance of the generator model in dealing with the mode collapse issue, the distributions of the real high-resolution HSI ($I^{HR}$) and the generated super-resolution HSI ($I^{SR}$) were visualised in the feature space, where the probability densities of the discriminator eigenvalues of $I^{HR}$ and $I^{SR}$, denoted as $D(I^{HR})$ and $D(I^{SR})$, were used to represent the sample distributions. With $I^{HR}$ and $I^{SR}$ as inputs separately fed into the discriminator $D$ described in Section \ref{sec:D}, the outputs from the last Maxpool layer, denoted as $D(I^{HR})$ and $D(I^{SR})$, represent the eigenvalues of the inputs $I^{HR}$ and $I^{SR}$ in the high-level discriminating space, and the probability densities of the $D(I^{HR})$ and $D(I^{SR})$ represent the sample distributions of $I^{HR}$ and $I^{SR}$. The coverage of probability densities between the $D(I^{HR})$ and $D(I^{SR})$ represent the mode similarity of the $I^{HR}$ and $I^{SR}$ to indicate whether model collapse occurs in the generator.\par
\begin{figure}[]   
	\centering  
	\includegraphics[width=3.6in]{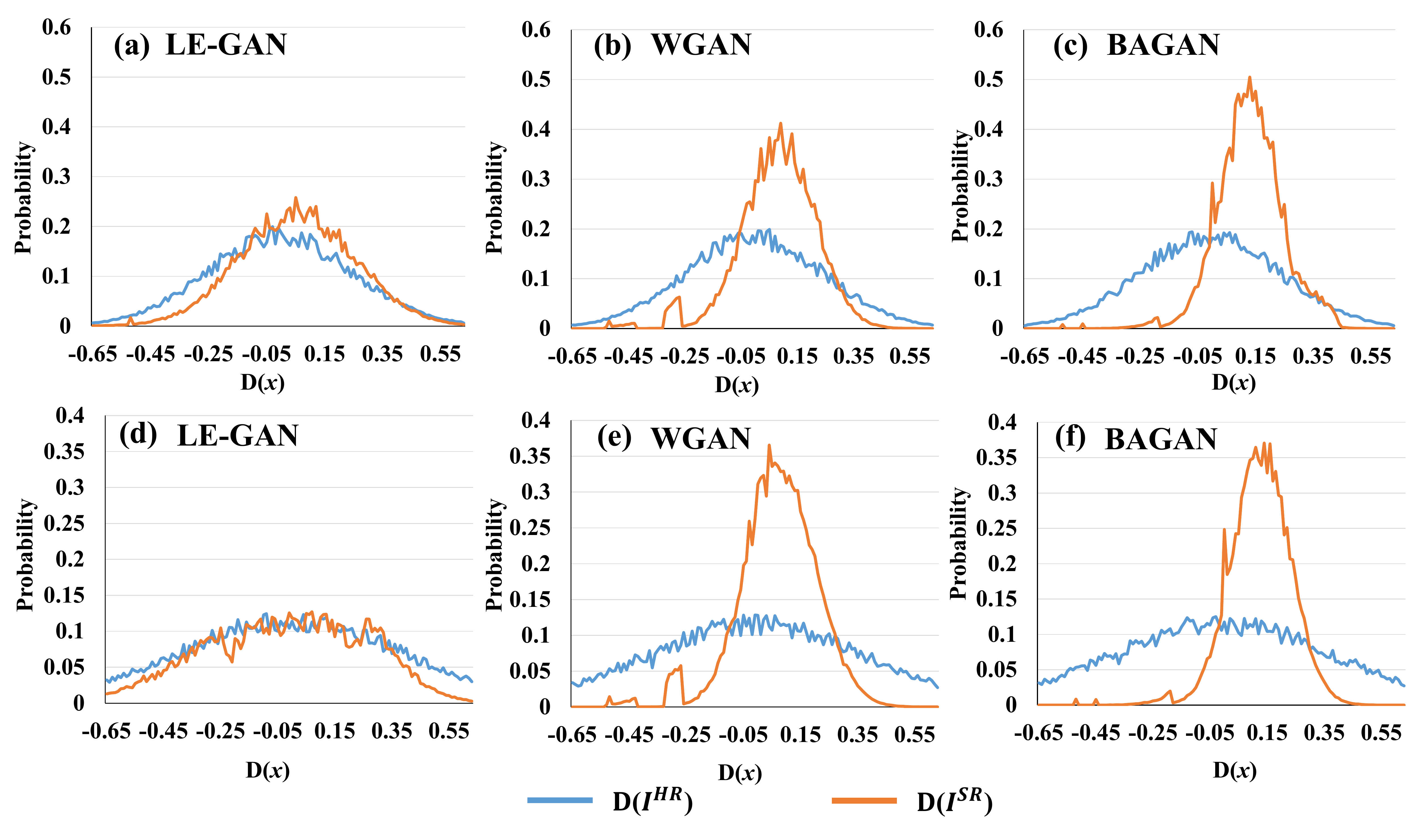}   
	\caption{Statistic distributions of the high-resolution HSI ($I^{hr}$) and the generated super-resolution HSI ($I^{SR}$) in the discriminator network ($D$). We tested the proposed LE-GAN with the two best competitors (i.e. WGAN and BAGAN) on AVIRIS (a-c) and UHD-186 (d-f) test dataset with an SNR level of 40db and an upscaling factor of $\times 8$. }    
	\label{fig:mode_collapse_2}  
\end{figure}

Fig. \ref{fig:mode_collapse_2} illustrates that the probability density curves of $D(I^{HR})$ and $D(I^{SR})$ obtained for three GAN models, LE-GAN and its two best competitors, WGAN and BAGAN, through training the models on AVIRIS and UHD-185 datasets with an SNR level of $40db$ and an upscaling factor of $\times 8$. In comparison with the other two models, the probability density curves of $I^{SR}$ generated by LE-GAN are much closer to those of the real $I^{HR}$ for both AVIRIS (Fig. \ref{fig:mode_collapse_2}a) and UHD-185 datasets (Fig. \ref{fig:mode_collapse_2}d). However, the probability density curves of the $I^{SR}$ generated by WGAN (\ref{fig:mode_collapse_2}b and e) and BAGAN (\ref{fig:mode_collapse_2}c and f) have an obvious tendency shifting towards the right and having a higher peak (i.e. a lower standard deviation). This means the $I^{SR}$ generated by WGAN or BAGAN can be better discriminated from the real $I^{HR}$ by $D$ (i.e. low spectral-spatial fidelity), and the generated $I^{SR}$ only covers the limited spectral-spatial patterns of the real $I^{HR}$ (i.e. existing the mode collapse issue). These observations shows that the proposed model outperforms the competitors in generating diversity of super-resolution samples and alleviating mode collapse.

\section{Discussion}

The challenge of GANs in improving the spectral and spatial fidelity of HSI super-resolution and addressing the issue of mode collapse is on how to make the generator learn the real spectral-spatial patterns, and meanwhile, prevent the generator from over-learning limited patterns. Since there is no such kinds of constraints in the JS distance based loss functions, the original GAN is hard to generate the high fidelity HSI super-resolution and easy to suffer mode collapse. In this study, we proposed a novel GAN model, named as LE-GAN, through improving the GAN baseline and introducing a new SSRP loss function. The new SSRP loss was used to guide the optimisation and alleviate the spectral-spatial mode collapse issue occurred in the HSI super-resolution process. The model validation and evaluation were conducted using the datasets from two hyperspectral sensors (i.e. AVIRIS and UHD-185) with various upscaling factors ($\times 2$, $\times 4$, and $\times 8$ ) and added-noises ($\infty$db, $40$db, and $80$db). The evaluation results showed that the proposed LE-GAN can achieve high-fidelity HSI super-resolution for relatively high upscaling factors and have a better robustness against noise and better generalizability to various sensors. 

\subsection{The ablation analysis of the improved modules}
In the proposed model, a total of five different modifications have been made to improve the GAN baseline including: 1) using 3D-convolutional filters in $G$, 2) adding an UpscaleBlock in $G$, 3) removing the sigmoid in $D$, 4) adding a novel $La$ network, and 5) using a new loss function to optimise the model.
 
To evaluate the effects of these improvements on the performance of the proposed LE-GAN, we have conducted an ablation analysis in which we gradually substituted the traditional GAN components with the proposed modules and compares their effects based on six evaluation metrics, PSNR, SIM, PI, SAM, SRE, and computing time (CT). Each improvement is an incremental modification to the original GAN model, thus forming five different models: Model 1 to Model 5. The details of the five models and their influences on the six evaluation metrics for the testing datasets (AVIRIS and UHD-185) with $\times 8$ scale factor are presented in Fig.\ref{fig:Model_assess}. The super-resolution results of three example patches are also displayed for the visual comparison. \par

\subsubsection{Model 1: using 3D-convolutional filters in $G$}
In order to process continuous spectral channels and capture spectral-spatial joint features learning in the ResBlock in $G$, 3D-convolutional filters are used. Theoretically, this modification is able to extract both the spectral correlation characteristics and spatial texture information. 

\subsubsection{Model 2: Adding an UpscaleBlock in $G$}
In a super-resolution network, the most important strategy to improve the performance is to increase the information (e.g. the dimensionality of feature maps) of an LR HSI to match with that of the corresponding HR HSI. However, the traditional approaches increase the feature dimensionality in the entire intermediate layers gradually, which increases the computation complexity and computational cost. In contrast, we proposed an UpscaleBlock to super-resolve the detailed spectral-spatial information only at the end of the generator network (see Fig. \ref{fig:G}). This adjustment directly eliminates the need of the computational and memory resources for super-resolution operations. Thus, a smaller filter size can be effectively used in our generator network for the extraction of super-resolution features. The results of Model 2 (the third column in Fig. \ref{fig:Model_assess}) reveals a performance improvement after adding the UpscaleBlock. Compared to Model 1, the computation time has a $35.1\%$ reduction on average without losing the super-resolution quality, Model 2 even has a better super-resolution quality in terms of PSNR, SSIM, PI, SAM and SRE.

\subsubsection{Model 3: Removing the sigmoid function from the discriminator}
In the traditional GAN framework, the sigmoid-activated features often skew the original feature distribution and result in lower reconstructed spectral values. Therefore, in this study, we removed the sigmoid activation in $D$ network for two reasons. Firstly, using the feature before activation can benefit accurate reconstruction of the spectral and spatial features of input. Secondly, the proposed latent space distance requires real feature distribution of the input HSI in the low-dimensional manifold in order to measure the divergence between the generated super-resolution HSI and real HR HIS. This modification, as shown in Model 3 in the fourth column of Fig. \ref{fig:Model_assess}, contributes to an approximately $7.2\%$ reduction in SAM and $13.9\%$ reduction in SRE. These findings suggest that removing the sigmoid activation can help keep the spectral consistency between the LR and HR HSIs.

\subsubsection{Model 4: Adding a newly developed La network}
The $La$ network is developed to produce a latent regularisation term, which holds up the manifold space of the generated super-resolution HSI so that the dimensionality of the generated HSI is consistent with that of real HR HIS. In addition, the $La$ network makes the divergence of the generated HSI and real HSI satisfy the Lipschitz condition for optimisation. The generated super-resolution HSI patches from Model 4 (see the fifth column of Fig. \ref{fig:Model_assess}) indicates that, after adding the $La$ network into the original GAN framework, both SAM and SRE have a significant reduction, with a drop of $6.7\%$ and $15.3\%$, respectively. Besides, there is a slight improvement on the PNSR, SSIM, and PI (the PNSR and SSIM respectively increase $2.8\%$ and $1.4\%$, the PI declines $4.3\%$). These results indicate that the regularisation term produced by the $La$ network has a great contribution in reconstructing the spectral-spatial details consistent with real HR HIS. However, the $La$ need to occupy a certain amount of computational and memory resources, subsequently the computation time increases $15.1\%$.

\subsubsection{Model 5: Using the new loss function to optimise the model}
\begin{figure}[]   
    \centering  
    \includegraphics[width=3.6in]{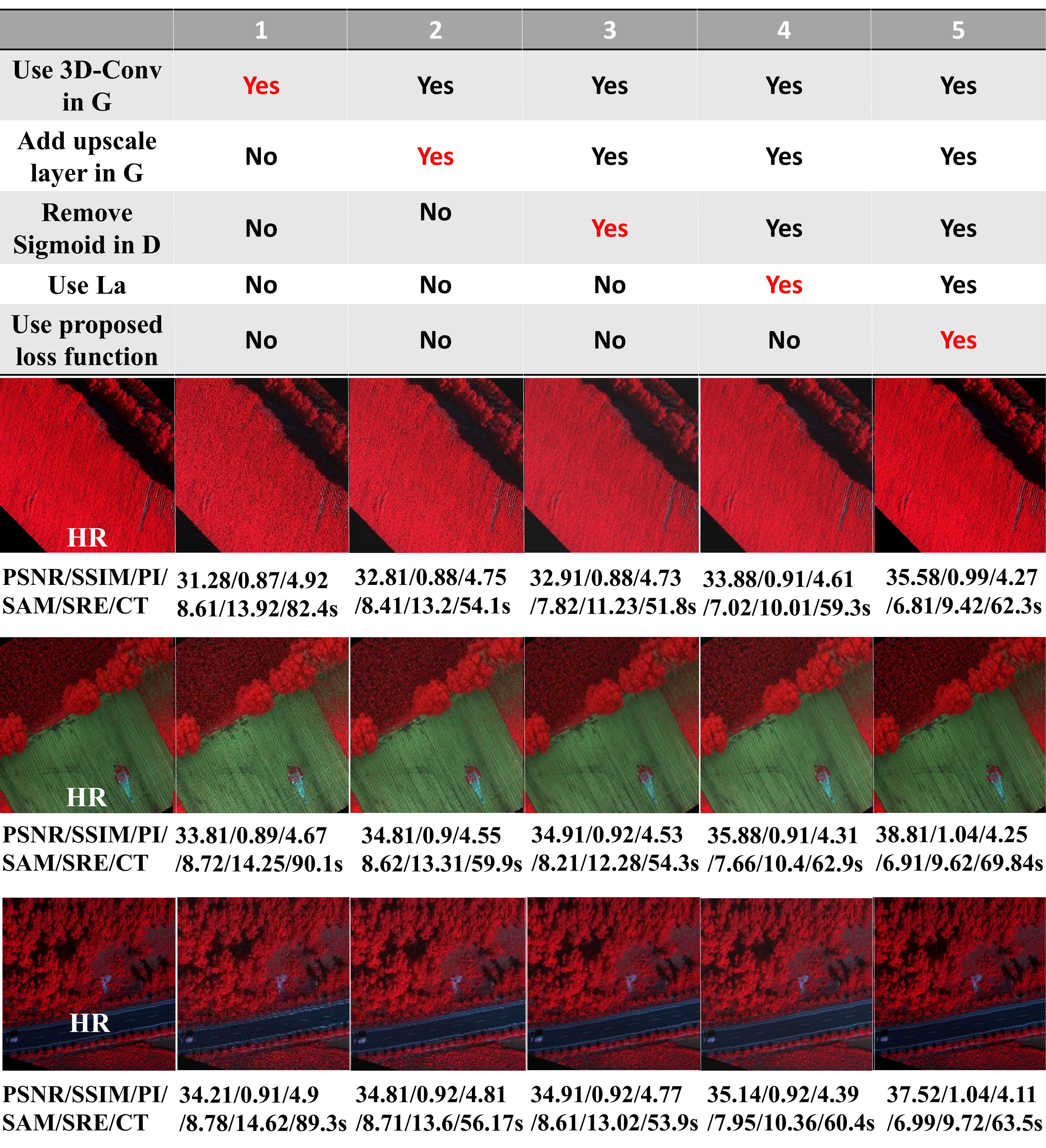}   
    \caption{The influence of five incremental modifications to the performance of the proposed model in terms of PSNR, SSIM, PI, SRE and computing time (CT). The results in each column (except first column) correspond to one new model with one incremental change to its previous model. The super-resolution results of different models with $\times 8$ scale factor on the testing datasets are illustrated for visual comparison.}    
    \label{fig:Model_assess}  
\end{figure}

The most important contribution of our work is to develop a SSRP loss function with a latent regularisation to optimise the whole model. Model 5 (see the last column of Fig. \ref{fig:Model_assess}), the final version of the LE-GAN, improves all of the evaluation metrics. The increases of PNSR and SSIM are $5.2\%$ and $12.4\%$,respectively, while the decreases of SAM and SRE are $13.1\%$ and $7.9\%$, respectively. But, it leads to a $11.4\%$ increase of computation time. These findings suggest that the proposed SSEP loss function with the latent space regularisation can boost the performance on measuring the divergence of generated HSI and real HSI in both spectral and spatial dimensionality.

\subsection{The Evaluation of the loss function}
\begin{figure}[]   
    \centering  
    \includegraphics[width=3.6in]{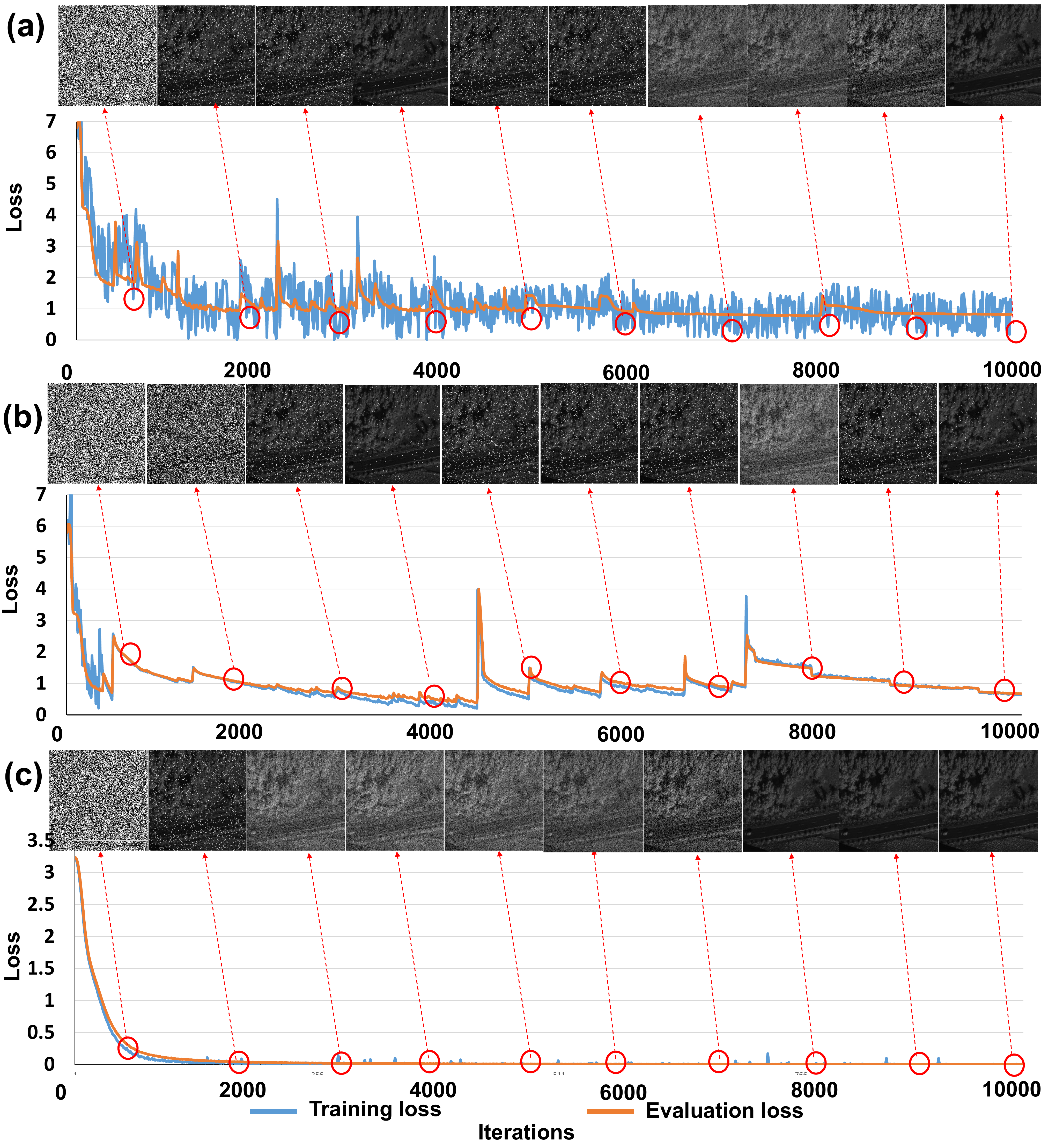}   
    \caption{A comparison of the loss curves during the training of Model 4 using (a) the traditional JS divergence-based loss (b) the Wasserstein distance-based loss, and (c) the proposed improved Wasserstein loss with latent regularisation.}    
    \label{fig:loss_val}  
\end{figure}

The proposed loss function introducing latent regularisation into the Wasserstein loss function optimises the GAN in the latent manifold space and addresses the problems of mode collapse. In order to verify the effectiveness of the proposed loss function, we trained the proposed LE-GAN model with three kinds of losses: 1) the traditional JS divergence-based loss, 2) the Wasserstein distance-based loss, and 3) the proposed improved Wasserstein loss with latent regularisation, and plotted their loss curves on both the training and validation sets in Fig. \ref{fig:loss_val}.\par

It is obvious that the training process of the model with a JS divergence-based loss, as shown in Fig. \ref{fig:loss_val}a, is unstable and volatile. The reason behind lies in the fact that the JS divergence always leads to the supports of $P_r$ and $P_g $ disjointing in the low-dimensional manifolds during the process of maximising the discriminative capability of $D$, which causes the gradient fluctuation. On the contrary, the Wasserstein distance based loss functions, as shown in Fig. \ref{fig:loss_val}b and c, can improve the stability of learning and lead the loss converges to the minimum. This findings is consistent with Arjovsky \textit{et al.} \cite{LN04}'s and Ishaan \textit{et al.} \cite{RN32}'s studies. In addition, it is noteworthy that the loss curve of the proposed model is more stable and smoother than that of the traditional Wasserstein distance-based losses. The theory behind is that introducing the latent regularisation terms into the training process provides a non-singular support to the generated sample sets at the corresponding low-dimensional manifolds. It is expected that the Wasserstein distance (i.e. $W(P_r,P_g)$) performs better under the condition of the continuity and differentiability of the divergence of $P_r $ and $P_g)$. With the latent regularisation, the max-min game of LE-GAN will yield a probability distribution $P_g (G(I^{lr}))$ in a low-dimensional manifold that has a joint support with $P_r (I^{hr})$, and the process of minimizing the $W(P_r,P_g)$ will facilitate the gradient descent of the trainable parameters in $G$ because the valid gradient can be captured from the optimised $D$ in the low dimensional manifold. Therefore, the latent regularisation derived Wasserstein loss is regarded as a more sensible loss function for HSI super-resolution than the JS divergence loss and the traditional Wasserstein loss. \par

The subplots above the learning curve shown in Fig. \ref{fig:loss_val} are the images generated in the optimisation process when three different losses are used. It is obvious that the super-resolved HSI subplots optimised by the JS divergence-based loss (see Fig. \ref{fig:loss_val}a) do not produce the equivalent quality of spatial texture reconstruction as those from the proposed model (see Fig. \ref{fig:loss_val}b). The proposed latent regularisation term makes the dimensionality of the generated HSI manifold more consistent with that of the HR HSI in the optimisation process.

\section{conclusion}
To address the challenge of spectral-spatial distortions caused by mode collapse during the optimisation process, this work has developed a latent encoder coupled GAN for spectral-spatial realistic HSI super-resolution. In the proposed GAN architecture, the generator is designed based on an STSSRW mechanism with a consideration of spectral-spatial hierarchical structures during the upscaling process. In addition, a latent regularised encoder is embedded in the GAN framework to map the generated spectral-spatial features into a latent manifold space and make the generator a better estimation of the local spectral-spatial invariances in the latent space. For the model optimisation, an SSRP loss has been introduced to avoid the spectral-spatial distortion in the super-resolution HSI. By using the SSRP loss, both spectral-spatial perceptual differences and adversarial loss in latent space are measured during the optimization process. More importantly, a latent regularisation component is coupled with the optimisation process to maintain the continuity and no-singularity of the generated spectral-spatial feature distribution in the latent space and increase the diversity of the super-resolution features. We have conducted different experimental evaluation in terms of mode collapse and performance. The proposed approach has been tested and validated on AVIRIS and UHD-185 HSI datasets and compared with five state-of-the-art super resolution methods. The results show that the proposed model outperforms the existing methods and is more robust to noise and less sensitive to the upscaling factor. The proposed model is capable of not only generating high quality super-resolution HSIs (both the spatial texture and spectral consistency) but also alleviating mode collapse issue.
\section*{Acknowledgment}

This research was supported BBSRC (BB/R019983/1), BBSRC (BB/S020969/1). The work is also supported by Newton Fund Institutional Links grant, ID 332438911, under the Newton-Ungku Omar Fund partnership (the grant is funded by the UK Department of Business, Energy, and Industrial Strategy (BEIS)) and the Open Research Fund of Key Laboratory of Digital Earth Science, Chinese Academy of Sciences(No.2019LDE003). For further information, please visit www.newtonfund.ac.uk.



%

%

\bibliographystyle{IEEEtran} 
\bibliography{Ref}



\end{document}